\documentclass [11pt]{article}
\pdfoutput=1 

\usepackage{graphicx}
\usepackage{amsmath,amssymb} 
\usepackage{latexsym}
\usepackage{mathrsfs}
\usepackage{bbold}
\usepackage{color}
\usepackage{slashed}
\definecolor{red}{rgb}{1,0,0}
\usepackage{cancel}
\usepackage{comment}
\usepackage{multirow}
\usepackage{arydshln}

\def\s{\!\!\!\!\textbf{/}}
\definecolor{red}{rgb}{1,0,0}
\makeatletter
\def\section{\@startsection {section}{1}{\z@}{-3.5ex plus -1ex minus
 -.2ex}{2.3ex plus .2ex}{\large\bf}}
\def\subsection{\@startsection{subsection}{2}{\z@}{-3.25ex plus -1ex
minus -.2ex}{1.5ex plus .2ex}{\normalsize\bf}}
\makeatother
\makeatletter

\@addtoreset{equation}{section}

\makeatother

\newcommand{\Tr}{\text{Tr}}

\newcommand{\SO}{\text{SO}}
\newcommand{\SU}{\text{SU}}
\newcommand{\U}{\text{U}}

\newcommand{\luv}[0]{\Lambda_{UV}}

\usepackage[colorlinks,linkcolor=blue,citecolor=blue,urlcolor=blue,linktocpage]{hyperref}
\newcommand{\hhref}[2][]{\href{http://arxiv.org/abs/#2#1}{arXiv:#2}}

\usepackage{tikz}
\usetikzlibrary{arrows,shapes}
\usetikzlibrary{trees}
\usetikzlibrary{matrix,arrows} 				
\usetikzlibrary{positioning}				
\usetikzlibrary{calc,through}				
\usetikzlibrary{decorations.pathreplacing}  
\usepackage{pgffor}							

\usetikzlibrary{decorations.pathmorphing}	
\usetikzlibrary{decorations.markings}
\tikzset{
    vector/.style={decorate, decoration={snake}, draw},
	provector/.style={decorate, decoration={snake,amplitude=2.5pt}, draw},
	antivector/.style={decorate, decoration={snake,amplitude=-2.5pt}, draw},
    fermion/.style={draw=black, postaction={decorate},
        decoration={markings,mark=at position 1 with {\arrow[draw=black]{>}}}},
    fermionbar/.style={draw=black, postaction={decorate},
        decoration={markings,mark=at position .55 with {\arrow[draw=black]{<}}}},
    fermionnoarrow/.style={draw=black},
    gluon/.style={decorate, draw=black,
        decoration={coil,amplitude=4pt, segment length=5pt}},
    scalar/.style={dashed,draw=black, postaction={decorate},
        decoration={markings,mark=at position .55 with {\arrow[draw=black]{>}}}},
    scalarbar/.style={dashed,draw=black, postaction={decorate},
        decoration={markings,mark=at position .55 with {\arrow[draw=black]{<}}}},
    scalarnoarrow/.style={dashed,draw=black},
    electron/.style={draw=black, postaction={decorate},
        decoration={markings,mark=at position .55 with {\arrow[draw=black]{>}}}},
	bigvector/.style={decorate, decoration={snake,amplitude=4pt}, draw},
}

\begin{document}

\begin{titlepage}
\begin{flushright}
LYCEN 2015-01
\end{flushright}
\vskip 1.0cm
\begin{center}
{\Large \bf Anarchic Yukawas and top partial compositeness:\\ the flavour of a successful marriage} \vskip 1.0cm
{\large Giacomo Cacciapaglia$^{a,b}$,\ Haiying Cai$^{a,b}$,\ 
Thomas Flacke$^c$,\ Seung J. Lee$^{c,d}$,\ Alberto Parolini$^{c,e}$,\ Hugo Ser\^odio$^c$} \\[1cm]
{\it $^a$ Universit\'e de Lyon, F-69622 Lyon, France; Universit\'e Lyon 1, Villeurbanne, France}\\[5mm]
{\it $^b$ CNRS/IN2P3, UMR5822, Institut de Physique Nucl\'eaire de Lyon F-69622 Villeurbanne Cedex,
France}\\[5mm]
{\it $^c$ Department of Physics, Korea Advanced Institute of Science and Technology, 335 Gwahak-ro, Yuseong-gu, Daejeon 305-701, Korea}\\[5mm]
{\it $^d$ School of Physics, Korea Institute for Advanced Study, Seoul 130-722, Korea}\\[5mm]
{\it $^e$ Center for Axion and Precision Physics, IBS, 291 Daehak-ro, Yuseong-gu, Daejeon 305-701, Korea}\\[5mm]
Email: \href{mailto:g.cacciapaglia@ipnl.in2p3.fr}{g.cacciapaglia@ipnl.in2p3.fr}, \href{mailto:hcai@ipnl.in2p3.fr}{hcai@ipnl.in2p3.fr}, \href{mailto:flacke@kaist.ac.kr}{flacke@kaist.ac.kr}, \href{mailto:sjjlee@kaist.ac.kr}{sjjlee@kaist.ac.kr}, \href{mailto:parolini@kaist.ac.kr}{parolini@kaist.ac.kr}, \href{mailto:hserodio@kaist.ac.kr}{hserodio@kaist.ac.kr}
\vskip 1.0cm%

\abstract{
The top quark can be naturally singled out from other fermions in the Standard Model due to its large mass, of the order of the electroweak scale.
We follow this reasoning in models of pseudo Nambu Goldstone Boson composite Higgs, which may derive from an underlying confining dynamics.
We consider a new class of flavour models, where the top quark obtains its mass via partial compositeness,
while the lighter fermions acquire their masses by a deformation of the dynamics generated at a high flavour scale.
One interesting feature of such scenario is that it can avoid all the flavour constraints without the need of flavour symmetries,
since the flavour scale can be pushed high enough. 
We show that both flavour conserving and violating constraints can be satisfied with top partial compositeness without invoking any flavour symmetry for the up-type sector,
in the case of the minimal \SO(5)/\SO(4) coset with top partners in the four-plet and singlet of \SO(4).
In the down-type sector, some degree of alignment is required if all down-type quarks are elementary. We show that taking the bottom quark partially composite provides a dynamical explanation for the hierarchy causing this alignment. We present explicit realisations of this mechanism which do not require to include additional bottom partner fields.
Finally, these conclusions are generalised to scenarios with non-minimal cosets and top partners in larger representations.}
\end{center}
\end{titlepage}

\newpage
\setcounter{tocdepth}{3}
\tableofcontents

\section{Introduction}

The discovery of a new scalar resonance at the LHC, resembling the Standard Model (SM) Higgs boson, unquestionably opened a new era in high energy particle physics. The Higgs boson is the last highly sought particle predicted by the SM, as originally proposed in the 70's. Nevertheless, the theoretical shortcomings of the SM, iconically represented by the naturalness problem, together with unexplained phenomena like Dark Matter and the baryon asymmetry in the Universe, cry for the presence of new physics. In some models, especially the ones addressing naturalness, the scale of new physics is close to the TeV, and possibly accessible at the LHC. The fact that no other state but a Higgs has been found in Run I of the LHC has not ruled out this possibility yet, as new states at the TeV scale, or lighter if weakly coupled, are still allowed.
On the other hand, the measurement of the couplings of the discovered Higgs boson offers a novel way to access the effects of new physics. At the moment, after the full analysis of the Run I data, the couplings are in good agreement with the SM predictions, at the 10\% level precision. Some couplings, like the ones to fermions, are much less constrained. Therefore, we are still in a situation where more fundamental realisations of the Higgs mechanisms other than the {\it ad-hoc} SM one are possible and deserve to be studied in depth, in view of the improvement in the Higgs coupling measurements expected at the Run II.

One very attractive idea, which dates back to the 70's~\cite{Weinberg:1975gm, Weinberg:1979bn, Susskind:1978ms}, is to replace the fundamental scalar field, responsible for the spontaneous breaking of the electroweak (EW) symmetry, with a whole strong interacting and confining sector. In this way, the scale at which the symmetry is broken is dynamically generated by a non-perturbative dimensional transmutation, which is thus insensitive to the problem of quantum instability. Furthermore, spontaneous  symmetry breaking by confinement is a very well known phenomenon, observed in many systems, notably in quantum chromodynamics (QCD). From the theoretical side, asymptotically-free and confining theories occupy the special seat of theories that can be potentially well defined at all scales.
The main obstacle of such idea is the difficulty in generating a large enough mass for the top quark and the absence of a light Higgs-like state.

In the 80's new ideas offered a revival of the whole scenario.
On the one hand, it was realised that, by extending the global flavour symmetries of the underlying dynamics, it is possible to generate a Higgs-like state among the pseudo-Nambu Goldstone Bosons (pNGB), like pions, of the spontaneous symmetry breaking. This mechanism allows a composite scalar to be naturally and parametrically lighter than other composite particles~\cite{Kaplan:1983fs,Kaplan:1983sm,Georgi:1984ef,Banks:1984gj,Georgi:1984af,Dugan:1984hq}.
In addition to it, for the top mass, the idea of partial compositeness \cite{Kaplan:1991dc} was put forward: in this scenario, the existence of fermionic spin-1/2 states with the same quantum numbers of the top is postulated. They couple to the source of electroweak symmetry breaking (EWSB), being part of the composite sector, while a linear coupling with the elementary fermions allows for a propagation of the symmetry breaking to the quark sector. The advantage over the traditional direct coupling of a top bilinear to the composite sector~\cite{Dimopoulos:1979es,Eichten:1979ah} is the absence of dangerous four-fermion operators that may mediate large flavour changing effects~\cite{Dimopoulos:1980fj}.

A third revival of the idea came in the new millennium when, inspired by the conjecture of a correspondence between warped extra dimensions (AdS) and conformal field theories in 4 dimensions (CFT) in string theory~\cite{Maldacena:1997re}, models of strong dynamics with a quasi-conformal behaviour have been associated to models on a warped extra dimensional background~\cite{Randall:1999ee, ArkaniHamed:2000ds}.
The extra dimensional version allows to build a weakly coupled model describing pions, together with spin-1 and spin-1/2 resonances, in a calculable way~\cite{Contino:2003ve}. On the other hand, the correspondence may not guarantee the existence of a 4-dimensional strongly coupled counterpart for any 5 dimensional model.
Following the success of extra dimensional models~\cite{Agashe:2004rs}, intensive studies of models of composite pNGB Higgs has sprouted, based on an effective Lagrangian approach (see Ref.~\cite{Bellazzini:2014yua} for a recent review).
Partial compositeness as the origin of fermion masses has been considered as a key ingredient at the basis of this kind of models. The main reason for this choice is that in the extra dimensional construction, flavour structures partially explaining the hierarchies in the fermion masses~\cite{ArkaniHamed:1999dc} and the absence of flavour changing neutral currents are automatically in place. Furthermore, loops of the composite fermions have been advocated as a stabilising mechanism for the Higgs mass, and the lightness of the Higgs mass has served as a motivation to consider fermions even lighter than the compositeness scale \cite{Matsedonskyi:2012ym,Marzocca:2012zn,Pomarol:2012qf}.
While this conclusion may be justifiable in models with a warped extra dimension where some of the top partners can be naturally lighter than the Kaluza-Klein scale~\cite{Agashe:2003zs, Carena:2006bn, Contino:2006qr}, in a 4D effective field theory approach the solidity of the assumption is more questionable, especially when all the top partners are assumed to be lighter than the compositeness scale. 
Examples of fundamental dynamics at the origin of this kind of models can be found, both non-supersymmetric~\cite{Galloway:2010bp} and supersymmetric~\cite{Caracciolo:2012je}.
More recently, the possibility to generate top partners in simple non-supersymmetric underlying dynamical models has been explored~\cite{Ferretti:2013kya}, with the conclusion that very few scenarios allow for such states to consistently exist~\cite{Barnard:2013zea,Ferretti:2014qta}. In fact, if the UV completion is a strong dynamics with only fermions as fundamental components, one is forced to add a number of fermions in different representations of the underlying hypercolour gauge group in order for the spectrum of composite states to contain spin-1/2 particles. Furthermore, additional multiplicity is implied by  QCD colour invariance which needs to be included within the global symmetries of the strong dynamics.

From the low energy effective field theory point of view, pNGB composite Higgs models (CHM) typically need flavour symmetries~\cite{Redi:2011zi} in order to satisfy the flavour bounds, as the flavour scale is set by the TeV scale for solving the naturalness problem. A comment is in order: in partial compositeness framework where all the quarks are partially composite, the flavour puzzle in new physics is mostly solved, a feature which makes partial compositeness very appealing. 
This is manifested in 5D models, where a Randall-Sundrum (RS) Glashow-Iliopoulos-Maiani mechanism is built-in~\cite{Agashe:2004cp}, representing a major achievement of partial compositeness. However, there is a residual tension left from $\epsilon_K$ and electric dipole moments (EDMs), which requires the compositeness scale to be still as high as $\mathcal{O}(10)$ TeV~\cite{Redi:2011zi,Delaunay:2010dw}. To solve this little hierarchy, some kind of flavour symmetries (e.g. horizontal symmetries, alignments, minimal flavor violation with $\SU(3)$ or $\U(2)$, etc ) are still required, both in 5D holographic models~\cite{Fitzpatrick:2007sa,Cacciapaglia:2007fw,Csaki:2008eh,Santiago:2008vq,Chen:2008qg,Csaki:2009wc} and 4D composite Higgs models~\cite{Redi:2011zi,Csaki:2008zd,Barbieri:2011ci,Barbieri:2012uh}. 

In this paper, we want to consider a scenario of pNGB composite Higgs where partial compositeness is present for the top quark only\footnote{Potentially also the bottom quark can be partially composite, {\emph{i.e.}} it is gaining its mass from linear mixing to composite fermions.}, while the masses of light quarks and leptons are generated by the more traditional direct coupling to the dynamics: an example of such structure could be found in technicolour theories~\cite{Dimopoulos:1979es,Eichten:1979ah}, in particular in conformal technicolour \cite{Luty:2004ye}. One reason for this choice is the difficulty in defining a simple underlying dynamics. Another more general issue relates to the properties of the dynamics that may be providing fermionic partners for each generation: in fact, if we assume that the strong dynamics respects a flavour symmetry and couples differently to the three generations, then it is inevitable to ask if such symmetry is spontaneously broken by the strong sector. Furthermore, including partners for all SM fermions would require a large number of fundamental fermions, that risks to spoil the asymptotic freedom of the underlying dynamics~\cite{Ferretti:2013kya}.
Without specifying a well-defined microscopic dynamics, it is not possible to answer these questions.
The case where both partial compositeness and direct couplings are present has some advantages: the top is uniquely defined as the combination of elementary fermions that couples to the fermionic partners, thus one can consider a truly flavour blind underlying dynamics.
Furthermore, as the large top mass is generated by the partial compositeness, the direct couplings can be suppressed by a larger energy scale (just light enough to generate the bottom, charm and tau masses) so that the sector responsible for the generation of such terms can be pushed to scales which are safe with respect to flavour violating effects, namely $O(10^{5})$ TeV. These benefits would however only hold if no additional flavour symmetries are needed.
We will therefore analyse the effects of bounds from both flavour-violating and flavour-conserving processes on a scenario where
the light fermion flavor structure is anarchic at the high flavour scale. 
We anticipate that two different small ratios play an important role: $m_c/m_t$ and $m_t/M_*$, where $M_*\sim1$ TeV indicates the scale of new physics. While the smallness of the latter is related to the smallness of $v/f$ and it is well understood in the context of pNGB Higgs models, $f\sim1$ TeV being the decay constant associated to the coset of the breaking, the former suppression is truly a result of the interplay of two different mass sources. The solution of the hierarchy problem needs to single out a combination of quarks and make it heavier, causing the emergence of an approximate $\U(2)$ symmetry in the up sector. A similar attitude has been also put forward in MSSM-like extensions of the SM, in particular in the context of deconstructed models \cite{Barbieri:2011ci,Craig:2011yk}.

It should be stressed that our scenario does not necessarily require the presence of a technicolour sector, nor of an underlying technicolour model. In fact, our analysis applies to a much larger class of models, including weakly coupled ones, described by the effective Lagrangian with a pNGB Higgs. Moreover models with heavy vector-like quarks, with or without a composite Higgs, might share the same properties as our framework.  

The paper is organised as follows: in Section~\ref{sec: MCHM5 plus TC} we introduce the minimal case of a SO(5)/SO(4) symmetry pattern with composite fermions in the {\bf 5} of SO(5), also known as MCHM$_5$, with only composite top and anarchic UV sector, and we present the constraints in this scenario in Section~\ref{sec:phenosec}, showing that the setup evades the need of any alignment or flavor symmetries in the top-sector and strongly reduces (but in general not completely overcomes) the need for alignment in the bottom-sector. In Section \ref{sec: bottom semiPC} we propose an extension in which the bottom quark mass is also realised  via partial compositeness which dynamically generates the residual alignment needed in the bottom sector. We present explicit realisations of this concept which do not require the inclusion of further composite partners. 
In Section~\ref{sec:general} we show that our results can be extended to non-minimal symmetry breaking pattern and also to cases where the top partners belong to other representations. Finally, we conclude in Section~\ref{sec:conclusion}.

\section{A composite Higgs model with additional Yukawa interactions}\label{sec: MCHM5 plus TC}

\begin{figure}[h]
\begin{tikzpicture}[line width=1.5 pt]
	\draw [line width=2 pt, ->] (-1,-0.13) -- (-1,1.5);
	\draw [line width=2 pt,] (-1,-5) -- (-1,-0.25);
	\draw (-1,-4.5) -- (-0.5,-4.5);
	\node at (0.8,-4.5) {EW scale $(v)$};
	\draw (-1,-3.5) -- (-0.5,-3.5);
	\node at (6.1,-3.5) {Compositeness scale $(f)$: the strong sector is described by heavy resonances,};	
	\node at (7.2,-4) {some of them have a mass of order $f$.};	
	\draw (-1,-1.5) -- (-0.5,-1.5);
	\node at (5.7,-1.5) {Condensation scale $(\Lambda_{HC})$: the strong dynamics breaks $\SO(5)$ to $\SO(4)$.};
	\draw (-1,1) -- (-0.5,1);
	\node at (5.1,1) {Flavour scale $(\luv)$: additional Yukawa operators are generated.};
	\draw[line width=1 pt] (-1.23,-0.25) -- (-0.75,-0);
	\draw[line width=1 pt] (-1.23,-0.37) -- (-0.75,-0.12);

	\draw[line width=1 pt, <->] (-2.2,-1.5) -- (-2.2,-3.5);
	\node[fill=white] at (-2.2,-2.25) {resonances};
	\node[fill=white] at (-2.2,-2.65) {tPC+HC};
 \end{tikzpicture}\caption{Schematic representation (not in scale) of the energy range considered in the model.} \label{fig:schema}
 \end{figure}
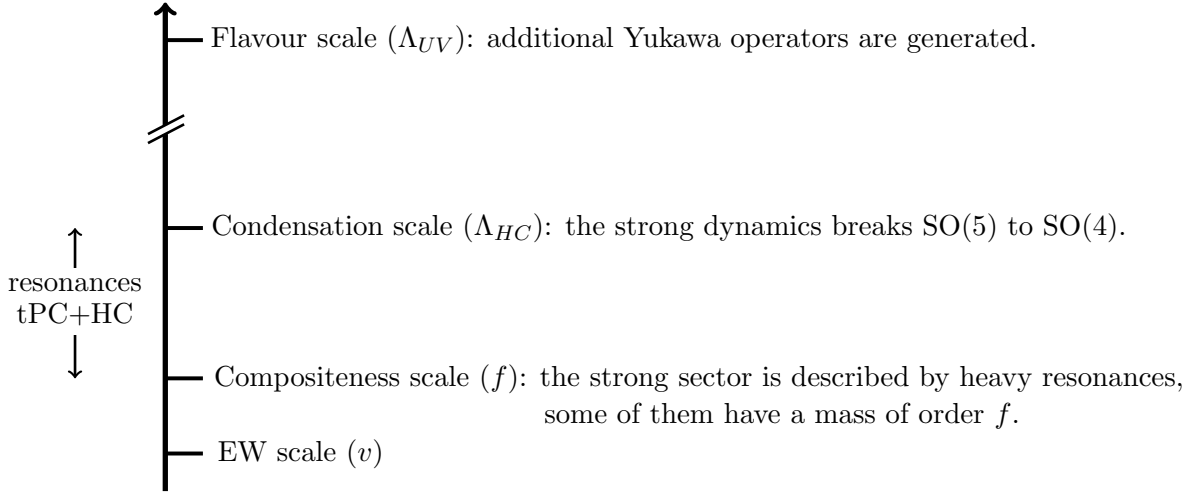

The basic set-up of the model can be represented by the diagram in Figure~\ref{fig:schema}: we imagine the existence of resonances of a strong dynamics sector appearing above a scale $f$, and deriving from the condensation of a ``hyper-colour" gauge group which takes place at a higher scale $\Lambda_{HC}$.
Between these two scales, all the heavy resonances appear, including top partners and spin-1 resonances. Above the scale $\Lambda_{HC}$, we postulate the existence of a different dynamics that generates direct couplings of the elementary fermions (both quarks and leptons) to the strong sector at a high scale: this may happen via four-fermion operators generated by a conformal technicolour sector (as an example).
In order to sufficiently suppress flavour changing neutral currents generated at the same scale, we will require that $\Lambda_{UV} \gtrsim 10^{5}$ without assuming CP conservation: this is in stark tension with the generation of the top mass, however such a  large scale can be compatible with the generation of the bottom and charm masses, as we will show in Section~\ref{sec:UVflavour}.
On the other hand, the compositeness scale $f$ needs to be at the TeV scale for consistency with naturalness.
Note also that we postulate the existence of a gap between $f$ and the EW scale $v$, so that the mass of the pNGBs of the symmetry breaking are at a scale $v$, compatible with the measured value of the Higgs mass.
The gap is also required for the model to pass electroweak precision tests, so that the confinement scale is larger than the EW scale, $f > v$, at the price of a fine tuning.

\subsection{Particle content and Lagrangian}
In this section we shall introduce a class of models which could give rise, at low energy, to the paradigm previously discussed. These are the well known CHM, where the Higgs boson arises as a pNGB of a spontaneously broken global symmetry, and the top quark couples to heavy fermions through partial compositeness, namely linear mass mixing. Both the spontaneous breaking, and the heavy fermions, may be produced by an underlying strong dynamics, however this is not the only scenario we want to cover. In this section, we restrict ourselves to the minimal coset preserving a custodial symmetry, namely $\SO(5)/\SO(4)$~\cite{Agashe:2004rs}
, and we embed the left-handed quarks $q_L={(t_L,b_L)}^T$ and the right-handed one $t_R$ in spurions transforming in the fundamental $\bf 5$ of $\SO(5)$ \cite{Contino:2006qr,Agashe:2006at}: this set-up is also known as MCHM$_5$. We also assign to the elementary spurions 
\begin{equation}\label{eq: SO5 spurions}
\overline{q}_{3L}^5=\frac{1}{\sqrt{2}}\left(-i\overline{b}_L,\, \overline{b}_L,\, -i\overline{t}_L,\, -\overline{t}_L,\,0\right)\,,\quad \overline{t}_R^5=\left(0,\,0,\,0,\,0,\,\overline{t}_R\right)\,,
\end{equation}
a charge $\pm2/3$ under an additional local ${\U(1)}_X$. 
The Higgs doublet, whose components are identified with the pNGBs, always appears through a unitary matrix $U$. In the unitary gauge, where the three real scalar degrees of freedom providing the longitudinal modes of the electroweak  $W^\pm$ and $Z^0$ gauge bosons disappear, it takes the form
\begin{equation}\label{eq: U}
U=
\begin{pmatrix}
1&0&0&0&0\\
0&1&0&0&0\\
0&0&1&0&0\\
0&0&0&c_\theta&s_\theta\\
0&0&0&-s_\theta&c_\theta
\end{pmatrix}\quad\text{with}\quad s_\theta=\sin\theta=\sin\frac{h+\left<h\right>}{f}\,,
\end{equation}
and the decay constant $f$ depends on the detail of the UV theory. Following naive arguments, we fix $4\pi f\simeq\Lambda_{HC}$. The pion matrix $U$ transforms non-linearly under ${\bf g}\in\SO(5)$: $U\rightarrow {\bf g}U{\bf h}^\dagger({\bf g},h)$, where ${\bf h}\in\SO(4)$. It is convenient to define $\Sigma=U\cdot {(0\,0\,0\,0\,1)}^t$, transforming linearly as a $\bf 5$ under $\SO(5)$. We also define
\begin{equation}
\epsilon=\langle h\rangle/f\,,
\end{equation}
$s_\epsilon=\sin\epsilon$ and $c_\epsilon=\cos \epsilon$. The EW scale is set by $v=s_\epsilon f\simeq246$ GeV, and we focus on values $s_\epsilon^2\sim 0.1$ which implies $f\simeq800$ GeV, as set by electroweak precision bounds~\cite{Agashe:2005dk}.

The composite sector typically contains many spin-1/2 fermionic resonances. We choose here to use the minimal set apt to generate a mass for the top via linear mixing, \emph{i.e.} a four-plet $Q$ and a singlet $\tilde{T}$ of $\SO(4)$, originating from a $\bf 5$ of $\SO(5)$:
\begin{equation}\label{def psi=4+1}
\psi=
\begin{pmatrix}
Q\\
\tilde{T}
\end{pmatrix}=\frac{1}{\sqrt{2}}
\begin{pmatrix}
i\ B-i\ X_{5/3}\\
B+X_{5/3}\\
i\ T+i\ X_{2/3}\\
-T+X_{2/3}\\
\sqrt{2}\ \tilde{T}
\end{pmatrix}\,.
\end{equation}
The most general Lagrangian we can write is then
\begin{align}\label{eq:Lag tot PC}
\begin{split}
\mathcal{L}_{comp}=&i\overline{Q}_{L,R}\left(D\s+E\s\right) Q_{L,R}+i\overline{\tilde{T}}_{L,R} D\s\tilde{T}_{L,R}-M_4\left(\overline{Q}_L Q_R+\overline{Q}_R Q_L\right)\\
&-M_1\left(\overline{\tilde{T}}_L\tilde{T}_R+\overline{\tilde{T}}_R\tilde{T}_L\right)+ic_L \overline{Q}_L^i\gamma^\mu d^i_\mu \tilde{T}_L+ic_R \overline{Q}_R^i\gamma^\mu d^i_\mu \tilde{T}_R+\text{h.c.}\\
-\mathcal{L}_{mix}=&y_{L4,1}f \overline{q}_{3L}^5 U\psi_R+y_{R4,1}f \overline{t}_R^5U\psi_L+\text{h.c.}\\
=&y_{L4}f \left(\overline{b}_L B_R+c^2_{\theta/2}\overline{t}_L T_R+s_{\theta/2}^2\overline{t}_L X_{2/3R}\right)-\frac{y_{L1}f}{\sqrt{2}}s_\theta \overline{t}_L\tilde{T}_R\\
&+y_{R4}f\left(\frac{s_\theta}{\sqrt{2}}\overline{t}_R T_L-\frac{s_\theta}{\sqrt{2}}\overline{t}_R X_{2/3L}\right)+y_{R1}f c_\theta \overline{t}_R \tilde{T}_L+h.c.\, ,
\end{split}
\end{align}
where $E_\mu$ and $d_\mu$ denote the CCWZ Cartan-Maurer one-forms ({\it c.f.}, {\it e.g.}, Ref.\cite{DeSimone:2012fs} for the explicit expressions).
Masses and couplings deriving from this Lagrangian are detailed in Appendix~\ref{secdetails}. The terms in $\mathcal{L}_{mix}$ are responsible for the partial compositeness of the top quark. Top partial compositeness and the gauging of ${\SU(2)}_L\times{\U(1)}_X$ lift the Higgs and radiatively induce EWSB. The detailed study of the Higgs potential is out of the scope of this work, it has been extensively investigated in \cite{Matsedonskyi:2012ym,Marzocca:2012zn,Pomarol:2012qf}, and we shall just rely on those results concerning the Higgs vacuum expectation value and mass. We fix the masses of the heavy coloured fermions appearing in Eq.(\ref{eq:Lag tot PC}) to be around $1$ TeV, and above the currents experimental bounds, roughly $650-800$ GeV depending on the quantum numbers and on the branching ratios \cite{Chatrchyan:2013uxa,Chatrchyan:2013wfa,Aad:2015kqa}.
We expect these fermionic composite objects to have masses of order $g_* f$ where $g_*$ is a coupling of the strong sector, implying masses in the few to multi-TeV range; on the other hand lighter masses are preferred by the light Higgs mass and therefore we assume this is the case and we loosely identify $M_*\sim M_4\sim M_1\sim |M_1-M_4|\sim f$. We do not expect heavier partners to invalidate the analysis we are about to develop. 
Note that Eq.(\ref{eq:Lag tot PC}) naturally singles out the top quark as the only elementary field that couples to the composite fermions.
We would also like to point out that, in the context of the model we are discussing, another possibility exists: to assume that $t_R$ is a fully composite state. With this choice, the Lagrangian giving rise to masses and couplings would be different from Eq.(\ref{eq:Lag tot PC})~\cite{Grojean:2013qca}\footnote{For other models with composite $t_R$, see~\cite{Panico:2012uw}.}. However, we have checked that the conclusions of the analysis we present here will stay the same, as the general flavour structure would be unaffected by changing $t_R$ from partially composite into fully composite (see Appendix~\ref{app:comptR}).

In addition we assume the presence of direct Yukawa interactions of all fermions, quarks and leptons, generated at a scale $\luv > \Lambda_{HC}$: they appear in the effective Lagrangian as couplings between pairs of SM elementary fermions and operators belonging to the composite sector. A simple example is provided by conformal technicolour-like theories, where they are four-fermion interactions with the component fermions of the strong dynamics. This mixed possibility, partial compositeness for the top and additional deformations for the other quarks, has been recently considered in \cite{Parolini:2014rza} in a supersymmetric theory and it has been analysed in the presence of flavour symmetries in \cite{Matsedonskyi:2014iha}. Similarly, it has been proposed in a non-supersymmetric model based on the coset SU(5)/SO(5)~\cite{Ferretti:2014qta}.  In this scenario we loose the nice feature of partial compositeness naturally generating flavour hierarchies but we can study microscopical models in realistic situations and still account for a single separation of scale, between the top and all the other quarks. Schematically, we complement the Lagrangian with the following interactions at the scale $\luv$
\begin{equation}\label{eq:4 ferm UV}
\mathcal{L}_Y= \bar{q}_{L,\alpha}\lambda^u_{\alpha,\beta}u_{R,\beta}\ \mathcal{O}_u +  \bar{\tilde{q}}_{L,\alpha} \lambda^d_{\alpha,\beta}d_{R\beta}\ \mathcal{O}_d+h.c.\,,
\end{equation}
where $\alpha$ and $\beta$ are indices over the 3 SM generations, and $\mathcal{O}_{u,d}$ are operators of the new dynamics. As these terms are generated independently on the partial compositeness of the top, their embedding in the global symmetry $\SO(5)$ is free. For concreteness, we will for now choose the same embedding as of the top, so that the spurions appearing in the above equations transform as $\bf 5$'s.
The fields $q_L$ and $u_R$ are thus a generalization of Eq.(\ref{eq: SO5 spurions}) to include three families, and $\tilde
{q}_L$ and $d_R$ are defined by
\begin{equation}\label{eq: down spurions}
\tilde{q}_{\alpha L}^5= 
\begin{pmatrix}
0&0&0&-1&0\\
0&0&1&0&0\\
0&1&0&0&0\\
-1&0&0&0&0\\
0&0&0&0&0
\end{pmatrix}
q_{\alpha L}^5\,,
\qquad
d_{\alpha R}^5=\left(\begin{array}{c}
0\\
0\\
0\\
0\\
d_{\alpha R}\end{array}\right)\,,
\end{equation}
with U$(1)_X$ charge $\pm1/3$. $\mathcal{O}_{u,d}$ are composite operators in a non trivial representation of the broken $\SO(5)$ interpolating at low energy the Higgs doublet. If these operators are in a representation contained in $\mathbf{5}\times\mathbf{5}$ of $\SO(5)$, at low energy we obtain the following 
\begin{align}\label{eq: 4 ferm Yukawa}
\begin{split}
\mathcal{L}_{Y}=&\sqrt{2}\ ({\bar{q}_{\alpha L}}^5\Sigma)m^u_{{\rm UV}\alpha\beta}(\Sigma^Tu_{\beta R}^5)+\sqrt{2}\ (\bar{\tilde{q}}_{\alpha L}^5\Sigma)m^d_{{\rm UV}\alpha\beta}(\Sigma^Td_{\beta R}^5)\\
=&\dfrac{s_{2\theta}}{2}\ \left[{\bar{u}_{\alpha L}}m^u_{{\rm UV}\alpha\beta}u_{\beta R}+{\bar{d}_{\alpha L}}m^d_{{\rm UV}\alpha\beta}d_{\beta R}\right]
\end{split}
\end{align}
where $m^{u,d}_{{\rm UV}}\propto\lambda^{u,d}$ such that $s_{2\epsilon}m^{u,d}_{\rm UV}\sim O(1)$ GeV to correctly reproduce the charm and bottom masses. The way $U$ appears is fixed by the representation of the operators $\mathcal{O}_{u,d}$: our choice gives the same dependence obtained for the top from partial compositeness\footnote{A simpler choice could be to have composite operators in the $\bf 5$ and embed right-handed quarks in $\SO(5)$ singlets: as a result we would have a different Higgs dependence in the effective Lagrangian, namely a single $\Sigma$ would appear. This choice would not significantly affect our analysis.}. If the composite sector is fundamentally a gauge theory of strongly interacting fermions ${\Psi}_i$ one can secretly imagine the operators $\mathcal{O}_{u,d}$ as ${\Psi}_i{\Psi}_j$ bilinears. If ${\Psi}$ transform in the $\bf 5$ of $\SO(5)$ at low energy we find the dependence expressed in Eq.(\ref{eq: 4 ferm Yukawa}). In this case in the UV the interactions, written in terms of microscopical degrees of freedom, are of the form $\bar{q}u{\Psi\Psi}/\Lambda_{\rm UV}^2$. This way of viewing them reminds us of conformal technicolour theories; in the following, we will dub these terms UV, independently on their physical origin.

These operators have a small impact on the Higgs potential: they do not play a significant role for what concerns naturalness. Indeed the largest contribution to the Higgs square mass is
\begin{equation}
-3 \frac{y_b^2}{16\pi^2}\Lambda_{HC}^2\simeq{(30\mbox{ GeV})}^2\,.
\end{equation}

Finally we stress that $\lambda^{u,d}$ are $3\times3$ generic matrices in generation space with rank $3$. This additional mass term of order $O(1\mbox{ GeV})$ in the up sector causes a misalignment between the physical top and the top defined as the partially composite quark.

\subsection{The structure of the model}
 
The fermionic field content defined above can be split into up and down sectors as
\begin{equation}
\xi_{\uparrow}=\begin{pmatrix}
u&c&t&T&X_{2/3}&\tilde{T}
\end{pmatrix}^T\,,\quad \xi_{\downarrow}=\begin{pmatrix}
d&s&b&B
\end{pmatrix}^T\,.
\end{equation}
Their Yukawa-mass Lagrangian is given by
\begin{align}
\begin{split}
-\mathcal{L}_{\rm yuk-mass}=&
{
\bar{\xi}_{\uparrow L}}\left[M_{\rm up}+Y_{\rm up}h+\cdots\right]
\xi_{\uparrow R}+
{
\bar{\xi}_{\downarrow L}}\left[M_{\rm down}+Y_{\rm down}h+\cdots\right]
{\xi}_{\downarrow R}+\text{h.c.}\\
\end{split}
\label{eq:xidef}
\end{align}
with the matrices $M_{\rm up}$ and $M_{\rm down}$ extracted from Eq.(\ref{eq:Lag tot PC}) and Eq.(\ref{eq: 4 ferm Yukawa}) and given in Appendix \ref{secdetails}. The first task is to define a proper change of basis in the up and down sector to recover the mass eigenstates. 

Since this cannot be done exactly we use $s_{2\epsilon}$ as an expansion parameter for perturbation theory: this implies that in the elementary quark sector a general $3\times 3$ matrix is a perturbation to the null matrix.  In other words, the unitary matrices that we shall find in this expansion do not completely diagonalize the $6\times 6$ (or $4\times 4$) matrix, but actually only block diagonalizes it. Nevertheless, this is enough since in this new basis the heavy eigenstates are diagonal and they can be safely integrated out at tree level.

For the up-quark sector we get, up to $\mathcal{O}(s_{2\epsilon}^3)$,
\begin{equation}\label{eq: diag up sec2}
U^{\dagger}_{uL} M_{\rm up} U_{uR}\simeq
\begin{pmatrix}
m_U&0\\
0&D_M
\end{pmatrix}\,,
\end{equation}
with
\begin{equation}\label{eq: muprime}
m_U\simeq \frac{s_{2\epsilon}}{2}\ m^u_{\rm UV}+m_t\Pi\,,\quad
\Pi=\small{\left(\begin{array}{ccc}0&0&0\\0&0&0\\0&0&1\end{array}\right)}\,,\quad
D_M\simeq\text{diag}\left(M_T,M_4,M_{\tilde{T}}\right)\,,
\end{equation}
where $m_t$ is the contribution to the top mass from partial compositeness (so that $ s_{2\epsilon} \ m_{\rm UV}\sim m_c \ll m_t$).
The masses are defined as
\begin{equation}\label{eq: masses block}
m_t=s_{2\epsilon}\frac{f^2|y_{L1}y_{R1}M_4-y_{L4}y_{R4}M_1|}{2\sqrt{2}M_T M_{\tilde{T}}}\,,
\,\, M_T=\sqrt{M_4^2+f^2 y^2_{L4}}\,,\,\, M_{\tilde{T}}=\sqrt{M_1^2+f^2 y^2_{R1}}\,.
\end{equation}
We also define
\begin{equation}
s_{\phi L}=\dfrac{y_{L4}f}{M_T}\,,\quad s_{\phi R}=\dfrac{y_{R1}f}{M_{\tilde{T}}}\,.
\end{equation}
The Yukawa couplings are brought, by these transformations, to a non block-diagonal form. The Yukawa matrix for the light quark sector is now given by
\begin{equation}\label{eq: Bu}
y_u\simeq \dfrac{m_U}{fs_{2\epsilon}/2}\left(1-\frac{1}{2}s_{2\epsilon}^2\right)+B_u
\,,\quad\mbox{where}\quad
B_u\sim\frac{\Sigma_u}{M_*^2}\,.
\end{equation}
We also define
\begin{equation}\label{eq: Sigma u}
\Sigma_u\sim\left(\begin{array}{ccc}
m_c^2&m_c^2&m_cm_t\\
m_c^2&m_c^2&m_cm_t\\
m_cm_t&m_cm_t&m_t^2
\end{array}\right)\,.
\end{equation}
Exact expressions are lengthy and are not reported here. They are obtained as outlined in Appendix \ref{sec:pertexp}. Here, we prefer to show approximate results capturing the size of the corrections. Hence these equations should not be considered as true equalities because we are neglecting numerical coefficients of order one.

For the down sector we obtain
\begin{equation}\label{eq:down masses}
U_{dL}^{\dagger} M_{\rm down}U_{dR}\simeq
\begin{pmatrix}
m_D&0\\
0&M_T
\end{pmatrix}\,,\quad m_D\simeq \frac{s_{2\epsilon}}{2}\ m^d_{\rm UV}\,.
\end{equation}
The Yukawa coupling in the down sector is decomposed in aligned and non aligned parts as  
\begin{equation}\label{eq:delta Yukawa down 5}
y_d\simeq\dfrac{m_D}{fs_{2\epsilon}/2}\left(1-\dfrac{s_{2\epsilon}^2}{2}\right)+B_d\,,
\end{equation} 
where in analogy with Eq.(\ref{eq: Sigma u}) we have
\begin{equation}
B_d\sim\frac{m_b\Sigma_d}{\epsilon M_*^3}\,,\quad\text{where}\quad \Sigma_d\sim\epsilon^2{(m^{d}_{\rm UV})}^2.
\end{equation}

The interaction Lagrangian of the EW gauge currents is
\begin{align}
\begin{split}
\mathcal{L}_{\rm gauge}=Z_\mu 
{
\bar{\xi}_{\uparrow L,R}}\gamma^\mu
A_{NC}^{tL,R}
\xi_{\uparrow L,R}+Z_\mu 
{
\bar{\xi}_{\downarrow L,R}}\gamma^\mu
A_{NC}^{bL,R}
\xi_{\downarrow L,R}+
W^+_\mu {
\bar{\xi}_{\uparrow L,R}
}\gamma^\mu
A_{CC}^{L,R}
\xi_{\downarrow L,R}+h.c.
\end{split}
\end{align}
where $A^{tL,R}_{NC}$, $A^{bL,R}_{NC}$ and $A^{L,R}_{CC}$ are reported in Eq.(\ref{eq: Z up MCHM}), Eq.(\ref{eq: Z down MCHM}) and Eq.(\ref{eq: W MCHM}). Applying the unitary transformations $U_{uL,uR}$ and $U_{dL,dR}$ to the EW gauge currents we obtain:
\begin{itemize}
\item deviations in the neutral currents
\begin{align}\label{eq:delta Z}
\begin{split}
\left.\delta A^{tL}_{NC}\right|_{3\times 3}\simeq& \dfrac{g}{c_W }\dfrac{\Sigma_u}{M_\ast^2}
\,,\quad
\left.\delta A^{tR}_{NC}\right|_{3\times 3}\simeq -\dfrac{g}{c_W}\dfrac{\Sigma_u}{M_\ast^2}
\,,\\
\left.\delta A_{NC}^{bL}\right|_{3\times 3}=&0\,,\quad
\left(\left.\delta A_{NC}^{bR}\right|_{3\times 3}\right)_{ij}\simeq
-\dfrac{g}{2 c_W }\frac{\Sigma_d}{M_\ast^2}\,;
\end{split}
\end{align}

\item deviations in the charged currents
\begin{align}\label{eq: delta charged}
\begin{split}
\left(\delta A_{CC}^{L}\right)\simeq&-\frac{g}{\sqrt{2}}\frac{\Sigma_u}{M_\ast^2}\,,\quad
\left(\delta A_{CC}^{R}\right)\simeq-\frac{g}{\sqrt{2} M_\ast^2}
m_b\left(\begin{array}{ccc}
m_c&m_c&m_c\\
m_c&m_c&m_c\\
m_t&m_t&m_t\,.
\end{array}\right)\,.
\end{split}\
\end{align}

\end{itemize}

In order to go from this basis to the ``true'' mass eigenbasis we just need to perform unitary transformations acting on the light sector only. The light mass matrices in Eq.(\ref{eq: muprime}) and Eq.(\ref{eq:down masses}) are diagonalized through unitary transformations as follows:
\begin{equation}\label{eq:V for masses}
m_U=V_{uL}M_UV_{uR}^\dagger\,,\quad m_D=V_{dL}M_DV_{dR}^\dagger
\end{equation}
where $M_U=\mbox{diag}(m_u,m_c,m_t)$ and $M_D=\mbox{diag}(m_d,m_s,m_b)$ are the masses of the six quarks. Given the fact that $\mathcal{O}(s_{2\epsilon}m_{\rm UV})\sim \mathcal{O}(m_c)$, the matrix $m_U$ given in Eq.(\ref{eq: muprime}) contains a strong hierarchy due to the $\{3,3\}$ entry which receives a contribution from partial compositeness of order $m_t\gg m_c$. Therefore it can be diagonalized through $V_{uL,R}$ of the form
\begin{equation}\label{eq: diagon masses}
V_{uL,R}\sim\left(\begin{array}{ccc}
O(1)&O(1)&O(\frac{m_c}{m_t})\\
O(1)&O(1)&O(\frac{m_c}{m_t})\\
O(\frac{m_c}{m_t})&O(\frac{m_c}{m_t})&1
\end{array}\right)\,.
\end{equation}
From this hierarchy and Eq.(\ref{eq: Sigma u}) we have 
\begin{equation}\label{eq: funny Sigma}
V_{uL}^\dagger\Sigma_u V_{uR}\sim\Sigma_u\,.
\end{equation}
Therefore deviations in the gauge couplings of up currents can be directly read from Eq.(\ref{eq:delta Z}).

In the down sector there is no \textit{a priori} hierarchy between the mass matrix entries, except from the fact that is has to accommodate the down-type quark spectrum.

\section{Confronting the model with data}\label{sec:phenosec}

In this section we confront our model with the present constraints coming from flavour conserving/violating processes and also comment on precision data, non linearities and neutron EDM. All these effects may have three distinct origins:
\begin{itemize}
\item[(1)] induced solely by the mixing effects due to top partial compositeness and direct Yukawa couplings, thus appearing as flavour-violating couplings of the $Z$, $W$ and Higgs;
\item[(2)] induced by heavy resonances, appearing at the compositeness scale;
\item[(3)] induced by the dynamics that generates elementary Yukawa couplings at the scale $\luv$.
\end{itemize}
The third type of effects will play no role in our framework, as we will show in Sec.~\ref{sec:UVflavour}. 

We can now proceed to evaluate the impact of the above results on SM measurements: we first discuss  flavour-conserving couplings, leaving flavour-violating effects for the following subsection.

\subsection{Flavour preserving processes}\label{subsection: delta g}

\subsubsection{Constraints from top partial compositeness}

We start with the tree level coupling of the top to $Z$ and $W$ boson, mainly affected by the partial compositeness mixings. The expressions we found for $\delta g_{Zt_L}$, $\delta g_{Wt_Lb_L}$ and $\delta g_{Zt_R}$ at $\lambda^u=\lambda^d=0$ agree with Eq.(6.6) and Eq.(6.7) of \cite{Grojean:2013qca}; we also checked that the following relation holds true:
\begin{equation}\label{eq: delta ZW}
\frac{\delta g_{Zt_L}}{g/c_W}=\frac{\delta g_{Wt_Lb_L}}{g/\sqrt{2}}\,,
\end{equation}
as expected \cite{delAguila:2000aa,delAguila:2000rc,Aguilar-Saavedra:2013pxa}. In the limit $y_{L1}=y_{L4}$, $y_{R1}=y_{R4}$ and $c_L=c_R=1/\sqrt{2}$, we obtain simple formulae that can be considered as an example of more general complicated expressions
\begin{equation}\label{eq: delta gZt}
\delta g_{Zt_L}\simeq-\frac{g}{c_W}{\left(\frac{m_{t}}{M_\ast}\right)}^2\frac{{(1-s_{\phi R}^2)}^2}{2s_{\phi R}^2}\,,\quad \delta g_{Zt_R}\simeq-\frac{g}{c_W}{\left(\frac{m_{t}}{M_\ast}\right)}^2\frac{(2-s_{\phi L}^2)}{2}\,.
\end{equation}
In the general case, corrections of the same order are obtained\footnote{Notice that $\delta g_{Zt_{L}}\rightarrow0$ if $y_{L4,L1}\rightarrow0$ and $\delta g_{Zt_R}\rightarrow 0$ if $y_{R4,R1}\rightarrow0$, a fact that might be obscured in Eq.(\ref{eq: delta gZt}).}.

The corrections to the $Z$ couplings to the top can be large, but no experimental bound on them is available.
Such deviation, however, also enters the coupling to charged currents: besides threatening the unitarity of the CKM matrix in the light flavours, as we will discuss in the next section, it affects the value of the coupling of the $W$ to third generation quarks.
The latter needs to be compatible with the direct measurement of $|V_{tb}|=1.021\pm0.032$ \cite{Agashe:2014kda}.
To satisfy the bounds, it is enough to have 
\begin{equation}
|{\delta A_{CC}^L}|^{1/2}\sim\left|\frac{m_{t}}{M_\ast}\frac{{(1-s^2_{\phi R})}}{\sqrt{2}s_{\phi R}}\right|\lesssim10^{-1}\,.
\end{equation}
At fixed $m_t$ this implies that $s_{\phi R}<1/2$ is disfavoured, unless we take $M_1$ to be much larger than $1$ TeV. 

For what concerns right-handed couplings, $\bar{t}_R\slashed{W}b_R$,  the expression in Eq.(\ref{eq: delta charged}) gives us a coefficient $\sim\frac{g}{\sqrt{2}}\frac{m_tm_b}{M_4^2}$: the same result holds in models with partially composite top and bottoms \cite{Vignaroli:2012si} and from the analysis presented there of $b\rightarrow s\gamma$ processes we read $M_4\gtrsim1$ TeV\footnote{We thank N. Vignaroli for a comment on this point.}.

For the couplings of the bottom quark we obtain
\begin{equation}
\delta g_{Zb_L}=0\,,\quad\delta g_{Zb_R}=-\frac{gs_{2\epsilon}^2}{8c_W}{\left(\frac{y_{L4}f m^d_{{\rm UV}33}}{M_4^2+y_{4L}^2f^2}\right)}^2 \simeq-\frac{g}{2c_W}s_{\phi L}^2 c_{\phi L}^2 {\left(\frac{m_{b}}{M_\ast}\right)}^2\,;
\end{equation}
deviations to the left-handed couplings vanish, as expected, because of the custodial symmetry~\cite{Contino:2006qr}, while corrections to the right-handed ones are sufficiently suppressed by the smallness of the bottom mass.

\subsubsection{Constraints from heavy resonances}

We now proceed inspecting subleading corrections along the line of recent works \cite{Grojean:2013qca,Matsedonskyi:2014iha}: those are especially important in the down sector, where the contribution of the compositeness is under control. At tree level corrections proportional to the momentum exchanged in the vertex are not forbidden by the custodial symmetry, and one can expect operators like
\begin{equation}\label{eq: Zbb q}
\mathcal{L}\sim s_{\phi L}^2\frac{\bar{b}\gamma_\mu D_\nu F^{\mu\nu}b}{m_V^2}\simeq s_{\phi L}^2{\left(\frac{m_Z}{m_V}\right)}^2\bar{b}\slashed{Z}b
\end{equation}
to arise, where $m_V$ is the mass of a heavy vector resonance. The coefficient of the operator in Eq.(\ref{eq: Zbb q}) is proportional to $s_{\phi L}^2$, the square of the mixing of $b_L$ with the $B_L$ top partner: this is the only effect since we do not have other partners and, therefore, we obtain corrections to $\delta g_{Zb_L}$ and not to $\delta g_{Zb_R}$.
At loop level a potentially sizable effect  comes from the presence of four top partners interactions, generated by exchanging a vector resonance at a scale $m_V$ defined as before. 
This will be proportional to $\log (m_V/M_4)$ and again to $s_{\phi L}^2$ since we need two mass insertions on the external legs to connect the vertex with two elementary $b_L$. For the same reason we do not expect a similar contribution arising for $\delta g_{Zb_R}$. 
Both tree level and loop corrections can be estimated to be around $10^{-3}$ for $m_V=3$ TeV and for sensible values of other parameters, thus satisfying the experimental bounds \cite{Batell:2012ca}.

Deviations for other quark couplings are suppressed by ${(m_c/M_*)}^2$ and ${(m_b/M_*)}^2$ for up and down type respectively and they are below the experimental bounds which can be found in \cite{ALEPH:2005ab} for charm and in \cite{Agashe:2014kda} for light quarks, the latter extracted from parity violation measurements in atomic physics.

\subsection{Flavour violating processes}\label{subsec: FCNC MCHM5}

Four-quark operators changing the flavour number by two units, {\emph{i.e.}} $|\Delta F|=2$ transitions, are common in beyond the SM scenarios and can place strong constraints on the new physics. They are typically described by an effective Lagrangian of the form
\begin{align}
\begin{split}
\mathcal{L}^{|\Delta F|=2}=&\sum_{i=1}^5C^{q_\alpha q_\beta}_i\mathcal{Q}_i^{q_\alpha q_\beta}+\sum_{i=1}^3\tilde{C}^{q_\alpha q_\beta}_i\tilde{\mathcal{Q}}_i^{q_\alpha q_\beta}
\end{split}
\end{align}
with the dimension six operators defined as
\begin{equation} \label{eq:dim6}
\begin{array}{rlrl}
\mathcal{Q}_1^{q_\alpha q_\beta }=&\left(\overline{q}_{\beta L}\gamma_\mu q_{\alpha L}\right)\left(\overline{q}_{\beta L}\gamma_\mu q_{\alpha L}\right)\,,&\tilde{\mathcal{Q}}_1^{q_\alpha q_\beta }=&\left(\overline{q}_{\beta R}\gamma_\mu q_{\alpha R}\right)\left(\overline{q}_{\beta R}\gamma_\mu q_{\alpha R}\right)\,,\\
\mathcal{Q}_2^{q_\alpha q_\beta }=&\left(\overline{q}_{\beta R} q_{\alpha L}\right)\left(\overline{q}_{\beta R} q_{\alpha L}\right)\,,&\tilde{\mathcal{Q}}_2^{q_\alpha q_\beta }=&\left(\overline{q}_{\beta L}  q_{\alpha R}\right)\left(\overline{q}_{\beta L} q_{\alpha R}\right)\,,\\
\mathcal{Q}_3^{q_\alpha q_\beta }=&\overline{q}_{\beta R}^a  q_{\alpha L}^b \overline{q}_{\beta R}^b  q_{\alpha L}^a \,,&\tilde{\mathcal{Q}}_3^{q_\alpha q_\beta }=&\overline{q}_{\beta L}^a  q_{\alpha R}^b \overline{q}_{\beta L}^b  q_{\alpha R}^a \,,\\
\mathcal{Q}_4^{q_\alpha q_\beta }=&\left(\overline{q}_{\beta R}q_{\alpha L}\right)\left(\overline{q}_{\beta L} q_{\alpha R}\right)\,,&&\\
\mathcal{Q}_5^{q_\alpha q_\beta }=&\overline{q}_{\beta R}^a  q_{\alpha L}^b  \overline{q}_{\beta L}^b  q_{\alpha R}^a \,.&&\\
\end{array}  
\end{equation}
In the down sector the most relevant constraints to these operators come from the $K^0-\overline{K}^0$ and $B_q^0-\overline{B}_q^0$ systems, described by the operators $\mathcal{Q}^{sd}$ and $\mathcal{Q}^{bq}$, respectively. In the up sector the $D^0-\overline{D}^0$ system place constraints on $\mathcal{Q}^{cu}$. We compute the coefficients $C_i$ at tree level and we compare with the bounds reported in \cite{Isidori:2010kg} for new physics scale at $1$ TeV. We neglect running effects, expected to introduce at most $O(1)$ variations.
We also discuss, when necessary, $|\Delta F|=1$ processes, such as strange meson decays, top flavour violating decays and non SM couplings of the $W$ boson. 

In the following we first survey the contributions from the three distinct sources (top compositeness, heavy resonances and UV operators), then we discuss the overall implications for our model. Finally, we close with a remark on neutron EDM.

\subsubsection{Constraints from top partial compositeness}

The first type of contributions will apply, mostly, to operators relevant to the  $D^0-\overline{D}^0$ system, due to the absence of bottom composite partners. Higgs flavour violating couplings are present in the theory and they are given by
\begin{equation}\label{eq: hup flav viol}
V_{u,L}^\dagger\,B_{u}V_{uR}\,,
\end{equation}
with $B_u$ given in Eq.(\ref{eq: Bu}).
The contribution of a Higgs exchange to the operator $\mathcal{Q}_4^{uc}$ can be estimated to be of the order of
\begin{equation}\label{eq: H deltaF up}
\frac{1}{m_H^2}{\left(\frac{m_c}{M_*}\right)}^4\simeq \frac{10^{-12}}{\mbox{TeV}^2} \left(\frac{1\ \mbox{TeV}}{M_\ast}\right)^4\,,
\end{equation}
where $M_*$ is a generic top-partner mass.
For what concerns the down sector we have negligible effects because $|B_d|\sim \epsilon (m_b/M_\ast)^3\sim10^{-7}(1\,\text{TeV}/M_\ast)^3$. 
Flavour violating $Z$ interactions are controlled by
\begin{equation}\label{eq: Z flav viol}
V_{uL}^\dagger\delta A^{tL}_{NC}V_{uL}\,,\quad V_{dL}^\dagger\delta A^{bL}_{NC}V_{dL}\,.
\end{equation}
where $\delta A^{t,bL}_{NC}$ are given in Eq.(\ref{eq:delta Z}). In this case the exchange of a $Z$ boson contributes to $\mathcal{Q}_1^{uc}$ with a coefficient proportional to
\begin{eqnarray}\label{eq:Z Q1 FCNC up}
{\left(V_{uL}^\dagger\delta A^{tL}_{NC}V_{uL}\right)}_{12,21}\quad\mbox{and given by}\quad\frac{g^2}{16c_W^2m_Z^2}{\left(\frac{m_c}{M_\ast}\right)}^4\simeq \frac{10^{-11}}{\mbox{TeV}^2} \left( \frac{1\ \mbox{TeV}}{M_\ast}\right)^4\,.
\end{eqnarray}
Therefore, flavour violation in the up sector is well under control.

In the down sector, the situation is different, since at tree level in our effective description $\delta A^{bL}_{NC}=0$. Therefore, we use here the contribution from higher order operators we discussed in Section \ref{subsection: delta g} for the $Zbb$ coupling; this results in effective operators of the form
\begin{eqnarray}\label{eq: down Z}
&&\frac{1}{m_Z^2}\left(s_{\phi L}\frac{m_Z}{m_V}\right)^4 \left[(V_{dL33}^\ast V_{dL31})^2\mathcal{Q}_1^{db}+(V_{dL33}^\ast V_{dL32})^2\mathcal{Q}_1^{sb}+(V_{dL32}^*V_{dL31})^2\mathcal{Q}_1^{ds}\right]\nonumber\\
&\simeq& \frac{10^{-4}}{\mbox{TeV}^2}\, 
\left[(V_{dL33}^\ast V_{dL31})^2\mathcal{Q}_1^{db}+(V_{dL33}^\ast V_{dL32})^2\mathcal{Q}_1^{sb}+(V_{dL32}^*V_{dL31})^2\mathcal{Q}_1^{ds}\right]\,,
\end{eqnarray}
for $m_V$ with a mass at 3 TeV. These coefficients are too large, therefore one need to rely either on the fact that the higher order operators are suppressed more than what naively expected, or the mixing angles in the down sector have a hierarchy. Comparing with Ref.\cite{Isidori:2010kg}, we find that
\begin{equation}\label{eq: down Z2}
|V_{dL33}^\ast V_{dL31}|<10^{-1}\,,\quad|V_{dL33}^\ast V_{dL32}|<10^{-1/2}\,,\quad |V_{dL32}^*V_{dL31}|<10^{-5/2}\,.
\end{equation}
These constraints are in mild tension with our assumption of anarchic masses, requiring some kind of alignment.

Flavour violating couplings of the $Z$ boson can also be constrained from $B_s\rightarrow\mu^+\mu^-$ decay branching ratios \cite{Guadagnoli:2013mru,Straub:2013zca}: from Eq.(\ref{eq:delta Z}) we easily read a suppression of the form $m_b^2/M_\ast^2$ whereas deviations up to order $10^{-3}$ are allowed.

Flavour violating neutral currents can also mediate flavour violating top decays, such as $t\rightarrow ch,uh$ and $t\rightarrow Zq$, which are being probed at the LHC \cite{Atwood:2013ica,Chatrchyan:2013nwa}.
In our framework we only have partial compositeness for the top quark, and therefore no flavour violation can arise from this sector alone. All flavour violation has to be linked with the flavour structure from the direct Yukawa couplings. As it can be see from Eq.(\ref{eq: Bu}) and Eq.(\ref{eq: hup flav viol}), the leading contributions to misaligned Yukawas are of the form 
\begin{equation}\label{eq:Higgs top FV}
y_{tc,L}\simeq y_{tc,R}\sim \dfrac{m_cm_t}{fM_*}\simeq10^{-4}\,.
\end{equation}
Third generation flavour violating $Z$ couplings are given in Eq.(\ref{eq:delta Z}) and Eq.(\ref{eq: Z flav viol}) and they read
\begin{equation}
 {(\delta A_{NC}^{tL,R})}_{32}\simeq\frac{g}{c_W}\frac{m_tm_c}{M_*^2}\simeq 10^{-4}\,.
\end{equation}
On the other hand we have \cite{Azatov:2014lha}
\begin{equation}
\mathcal{B}(t\rightarrow ch)\simeq0.25\ ({|y_{tc,L}|}^2+{|y_{tc,R}|}^2)\,,\quad\mathcal{B}(t\rightarrow Zc)\simeq3.5 \ {(\delta A_{NC}^{tL,R})}_{32}^2\,.
\end{equation}
Therefore, the effects expected in our scenario are many orders of magnitude too small to be detected at the LHC. In fact, after Run I,  $\mathcal{B}(t\rightarrow ch)<6\div8\times10^{-3}$ at 95 $\%$ CL \cite{Aad:2014dya,CMS:2014qxa} and $\mathcal{B}(t\rightarrow Zc)<5\times10^{-4}$ at 95 $\%$ CL, a limit set by CMS with $19.7$ fb$^{-1}$ at $\sqrt{s}=8$ TeV \cite{Chatrchyan:2013nwa}. During Run II, the LHC is expected to set limits up to $\mathcal{B}(t\rightarrow Zq)\sim O(10^{-4\div5})$ with $300$ and $3000$ fb$^{-1}$ \cite{CMS:2013zfa}, and similarly 
an improvement of order one on the bound on the Yukawa flavour violating couplings is expected with 300 fb$^{-1}$ data \cite{Atwood:2013ica}.
The results on Eq.~\eqref{eq:Higgs top FV} are in contrast with the usual case of partially composite light quarks, where the light effective Yukawa couplings are aligned with the mass matrix, resulting in the absence of flavour violation at $\mathcal{O}(y_Ly_R/M^2)$~\cite{Agashe:2009di}. It is then common to consider higher order contributions in the kinetic terms of the elementary quarks in order to estimate the dominant effects in Higgs FCNCs processes. In~\cite{Azatov:2014lha} the authors estimated these contributions through the help of holographic techniques and found, for the anarchic scenario, at $\mathcal{O}(y_L^2y_R^2/M_\ast^4)$
\begin{equation}
y_{tc,L}\sim \dfrac{m_tm_c}{fM_\ast V_{cb}}\,,\quad y_{tc,R}\sim \dfrac{m_t^2 V_{cb}}{fM_\ast }
\end{equation}  
in the quark mass eigenbasis. This type of flavour misalignment has been also estimated in~\cite{Agashe:2009di} through the use of naive dimensional analysis, and in~\cite{Azatov:2009na} in a specific 5D implementation using the mass insertion approximation in KK language.

Turning to $W$ couplings, in this model the CKM matrix is not a unitary matrix due to the presence of 3 extra tops and 1 extra bottom. It is defined by the following expression:
\begin{equation}\label{eq: CKM}
V_{CKM}={V_{uL}^\dagger(\mathbb{1}+\frac{\sqrt{2}}{g}\delta A^{L}_{CC}) V_{dL}}\,.
\end{equation}
As the matrices  $V_{uL,dL}$ are unitary, the corrections $\delta A^{L}_{CC}$ is constrained by unitarity, in particular 
\begin{eqnarray}\label{eq: CKM uni}
V_{uL}^\dagger V_{dL}&=&V_{CKM}-\frac{\sqrt{2}}{g} V_{uL}^\dagger \delta A^{L}_{CC}\, V_{dL}\,,\\
{\left(V_{uL}^\dagger V_{dL}\right)}^\dagger \left(V_{uL}^\dagger V_{dL}\right)&=&\mathbb{1}\quad\Rightarrow\quad \frac{\sqrt{2}}{g} V_{dL}^\dagger(\delta A^{L}_{CC}+\delta A^{L\dagger}_{CC})V_{dL}=V_{CKM}^\dagger V_{CKM}-\mathbb{1} + \mathcal{O}(\delta^2 A^{L}_{CC})\,,\nonumber\\
\left(V_{uL}^\dagger V_{dL}\right) \left(V_{uL}^\dagger V_{dL}\right)^\dagger&=&\mathbb{1}\quad\Rightarrow\quad \frac{\sqrt{2}}{g}V_{uL}^\dagger(\delta A^{L}_{CC}+\delta A^{L\dagger}_{CC})V_{uL}=V_{CKM}V_{CKM}^\dagger-\mathbb{1}+ \mathcal{O}(\delta^2 A^{L}_{CC})\,.\nonumber
\end{eqnarray}
Because of the unitarity of $V_{uL,dL}$, even taking small mixing angles, the unitarity of the CKM matrix cannot be restored, and the largest deviation appears in the coupling of the top to the bottom, which can easily be set to satisfy the bounds as we have seen in the previous section.
The unitarity violation in the up sector is also under control, thanks to the hierarchy in the 3$\times$3 mixing angles of order $m_c/m_t$. However, in the down sector, a hierarchy in the UV masses is also required in order to satisfy the bounds: comparing with the experimental results~\cite{Agashe:2014kda}, we find
\begin{equation}
{|V_{dL13}|}<10^{-1}\,,\quad{|V_{dL23}|}^2<10^{-1}\,.
\end{equation}
Once more, this indicates a mild hierarchy in the down-type sector, requiring some sort of alignment.

\subsubsection{Constraints from heavy resonances}
The effects induced by heavy resonances, in a way, reflect our little knowledge about the physics at the compositeness scale. One expects other resonances (vector, scalar, etc.) to appear at this scale, whose presence affects flavour observables. In a bottom up approach we shall parametrise scalar and vector resonances and look at the predictions for the $d=6$ operator coefficients. We will assume that the composite resonances only couple to composite fermions, even though direct couplings to the elementary quarks may be generated by the same mechanism coupling them to the strong sector to give them masses. Let us consider first the interaction of a real scalar field $\Phi$ transforming as a singlet of \SO(4):
\begin{equation}\label{eq: Phi interaction}
\mathcal{L}=\Phi(g_B \bar{Q} Q+ g_S \bar{\tilde T} \tilde{T})+\frac{1}{2}m_\Phi^2\Phi^2\,.
\end{equation}
The mass of this additional resonance is proportional to $f$ and to some strong coupling constant of the theory and it is expected to lie between $f$ and $\Lambda_{HC}$. Due to partial compositeness, Eq.(\ref{eq: Phi interaction}) induces interactions between the top and this additional scalar and, after diagonalising the quark mass matrices, this results in flavour violating couplings. Their flavour structure is
\begin{equation}\label{eq:Yukawa Phi up}
\mathcal{L}\simeq
\Phi\left(\begin{array}{ccc}\bar{u}_L&\bar{c}_L&\bar{t}_L\end{array}\right)\cdot\left(\begin{array}{ccc}
0&0&g_S s_{\phi R}^2c_{\phi R}\frac{m_c}{M_*}\\
0&0&g_S s_{\phi R}^2c_{\phi R}\frac{m_c}{M_*}\\
g_B s_{\phi L}^2c_{\phi L}\frac{m_c}{M_*}&g_Bs_{\phi L}^2c_{\phi L}\frac{m_c}{M_*}&(g_B-g_S)\frac{m_t}{M_\ast}
\end{array}\right)\cdot
\left(\begin{array}{c}u_R\\c_R\\t_R\end{array}\right)\,+h.c.
\end{equation}
In the mass eigenstates basis for the quarks the flavour violating vertex has thus the form
\begin{equation}
\mathcal{L}\simeq \tilde{g}{\left(\frac{m_c}{m_t}\right)}^2\frac{m_t}{M_*}\,\Phi\, \bar{u}c + h.c.\,,
\end{equation}
with $\tilde{g}\sim g_{B,S}$.  Integrating out $\Phi$ allows us to compute the coefficients of the dimension-6 operators in Eq.(\ref{eq:dim6}). In the case at hand, we are left with:
\begin{equation}\label{eq: coeff FV up Phi}
\mathcal{L}\simeq\left(\frac{\tilde{g}}{m_\Phi}\right)^2{\left(\frac{m_c}{m_t}\right)}^4{\left(\frac{m_t}{M_*}\right)}^2\mathcal{Q}_4^{uc}\simeq \left( \frac{1\ \mbox{TeV}}{M_\ast} \right)^2\left(\frac{\tilde{g}}{m_\Phi/\mbox{TeV}}\right)^2 \times \frac{10^{-10}}{\mbox{TeV}^{2}}\,\mathcal{Q}_4^{uc}\,,
\end{equation}
potentially larger than the effect of a misaligned Higgs Yukawa in Eq.(\ref{eq: H deltaF up}), but still well below the experimental bound \cite{Isidori:2010kg}.
For the down sector the induced Yukawas are only proportional to $g_B$ because we do not have partial compositeness for the right bottom. The analogous of Eq.(\ref{eq:Yukawa Phi up}) is
\begin{equation}\label{eq: Phi down}
\mathcal{L}\simeq g_B s_{\phi L}^2c_{\phi L}\,  \Phi\bar{b}_L\left(\frac{m_b}{M_*} d_R+\frac{m _b}{M_*}s_R+\frac{m_b}{M_*}b_R\right)+h.c.\,.
\end{equation}
Then, after going to the mass basis and integrating out the scalar resonance we get, for the flavour violating operator in the down sector analogous to Eq.(\ref{eq: coeff FV up Phi}),
\begin{align}\label{eq: down Phi}
\begin{split}
\mathcal{L}\simeq&\left(\frac{g_Bs_{\phi L}^2c_{\phi L}}{m_\Phi}\right)^2{\left(\frac{m_b}{M_*}\right)}^2\left[z_4^{db}\mathcal{Q}_4^{db}+z_4^{sb}\mathcal{Q}_4^{sb}+z_4^{ds}\mathcal{Q}_4^{ds}\right]\\
\simeq&  \left(\frac{1\,\mbox{TeV}}{M_\ast}\right)^2\left(\frac{g_B}{m_\Phi/\mbox{TeV}} \right)^2\times \frac{10^{-5}}{\mbox{ TeV}^{2}}\,\left[z_4^{db}\mathcal{Q}_4^{db}+z_4^{sb}\mathcal{Q}_4^{sb}+z_4^{ds}\mathcal{Q}_4^{ds}\right]
\end{split}
\end{align}
with the dimensionless coefficients given by
\begin{equation}
z_4^{d_\alpha d_\beta}=V_{dL3\alpha}^*V_{dL3\beta}\sum_{\gamma\delta} V_{dR\gamma\beta}V_{dR\delta\alpha}^\ast\,.
\end{equation}
The constraints on the $\mathcal{Q}_4$ operators require the dimensionless coefficients to satisfy
\begin{equation}\label{eq: phi down z bound}
|z_4^{db}|<10^{-2}\,,\quad|z_4^{sb}|^2<10^{-1}\,\,,\quad |z_4^{ds}|<10^{-6}\,\,,
\end{equation}
The constraints above can be considered as a conservative worse case scenario, as they have been computed by assuming $m_\Phi/g \sim f \sim 1$ TeV. In fact, the resonances may have a larger mass, up to the condensation scale $\Lambda_{HC} \sim 4 \pi f$ and have couplings to top partners of order one. Depending on the details of the underlying theory, therefore, the bounds may be mitigated by a factor up to $(4\pi)^2 \sim 10^2$.

In the case of massive vector resonances we can write the interaction in the form
\begin{equation}\label{eq: V interaction}
\mathcal{L}=V_\mu(g_B \bar{Q}_{L}\gamma^\mu Q_{L}+ g_S \bar{\tilde{T}}_{L}\gamma^\mu\tilde{T}_{L})+(L\rightarrow R)+\frac{1}{2}m_V^2V_\mu V^\mu\,.
\end{equation}
After bringing the mass matrices to their block diagonal form, the resonant vector contribution becomes
\begin{equation}
\mathcal{L}= V_\mu \bar{u}_{\alpha L,R}\gamma^\mu  (\delta A_{LR,res}^u)_{\alpha\beta} u_{\beta L,R}+
V_\mu \bar{d}_{\alpha L,R}\gamma^\mu  (\delta A_{LR,res}^d)_{\alpha\beta} d_{\beta L,R}
\end{equation}
with
\begin{eqnarray}\label{eq:delatAres}
\delta A_{L,res}^u\sim\left(\begin{array}{ccc}
g_Ss^2_{\phi R} c^2_{\phi R}\frac{m_c^2}{M_\ast^2}&g_Ss^2_{2\phi R}c^2_{\phi R}\frac{m_c^2}{M_\ast^2}& -(g_S-g_Bc_{\phi L}^2)c_{\phi R}\frac{m_t m_c}{M_*^2}\\g_Ss^2_{\phi R}c^2_{\phi R}\frac{m_c^2}{M_\ast^2}&g_Ss^2_{\phi R}c^2_{\phi R}\frac{m_c^2}{M_\ast^2}& -(g_S-g_Bc_{\phi L}^2)c_{\phi R}\frac{m_t m_c}{M_*^2}\\
 -(g_S-g_Bc_{\phi L}^2)c_{\phi R}\frac{m_t m_c}{M_*^2}&  -(g_S-g_Bc_{\phi L}^2)c_{\phi R}\frac{m_t m_c}{M_*^2}& g_B  s_{\phi L}^2+\frac{g_S-g_B}{s_{\phi R}^2}\frac{m_t^2}{M_\ast^2}
\end{array}\right)\,,
 \end{eqnarray}
where the right-handed couplings can be obtained from the above expression with the replacements $\phi_L\leftrightarrow \phi _R$ and $g_B\leftrightarrow g_S$, and
\begin{equation}
\delta A_{L,res}^d\sim
g_Bs_{\phi L}^2\left(
\begin{array}{ccc}
0&0&c_{\phi L}^3\frac{m_b^2}{M_\ast^2}\\
0&0&c_{\phi L}^3\frac{m_b^2}{M_\ast^2}\\
c_{\phi L}^3\frac{m_b^2}{M_\ast^2}&c_{\phi L}^3\frac{m_b^2}{M_\ast^2}&1+2c_{\phi L}^4\frac{m_b^2}{M_\ast^2}
\end{array}
\right)\,,\quad \delta A_{R,res}^d\sim g_B s_{\phi L}^2 c^2_{\phi L}\frac{\Sigma_d}{M_\ast^2}\,.
\end{equation}
In the mass eigenstates basis for the quarks, the coefficients of the flavour violating effective operators induced by the decoupling of the heavy resonance are 
\begin{align}\label{eq:heavy vectors C}
C_1^{uc}\simeq&  \frac{1}{m_V^2}\left( g_Bs_{\phi L}^2\left(\frac{m_c}{m_t}\right)^2+g_\ast^\prime\left(\frac{m_c}{M_\ast}\right)^2\right)^2
\sim \left(\frac{g_B}{m_V/\mbox{TeV}}\right)^2\times \frac{10^{-9}}{\mbox{ TeV}^{2}}\\
C_1^{d_\alpha d_\beta}\sim&\left(\frac{g_Bs_{\phi L}^2}{m_V}\right)^2\left((V^\dagger_{dL}\Pi V_{dL})_{\alpha\beta}+c_{\phi L}^3\frac{(V^\dagger_{dL}\Sigma_dV_{dL})_{\alpha\beta}}{M_\ast^2}\right)^2
\sim\left(\frac{g_B}{m_V/\mbox{TeV}}\right)^2\times \frac{\left[V_{dL3\alpha}^\ast V_{dL3\beta}\right]^2}{\mbox{TeV}^{2}}\nonumber\\
\tilde{C}_1^{d_\alpha d_\beta}\simeq&\left(\frac{g_B s_{\phi L}^2 c^2_{\phi L}}{m_V}\right)\left[\frac{(V^\dagger_{dR}\Sigma_dV_{dR})_{\alpha\beta}}{M_\ast^2}\right]^2\leq\left( \frac{1\ \mbox{TeV}}{M_\ast} \right)^4\left(\frac{g_B}{m_V/\mbox{TeV}}\right)^2\times \frac{10^{-10}}{\mbox{ TeV}^{2}}\nonumber
\end{align}
with $g_\ast^\prime\sim g_{S,B}$. The coefficient $\tilde{C}_1^{uc}$ can be obtained from $C_1^{uc}$ using the replacement described below Eq.\eqref{eq:delatAres}. The constraints on $C_1^{bd}$, $C_1^{bs}$ and $C_1^{sd}$ imply, respectively,
\begin{equation}\label{eq: heavy vector}
|V_{dL33}^\ast V_{dL31}|<10^{-3}\,,\quad |V_{dL33}^\ast V_{dL32}|<10^{-2}\,,\quad |V_{dL32}^\ast V_{dL31}|<10^{-5}\,.
\end{equation}
Similarly to the case of scalar resonances, a larger mass and a smaller coupling may lift the bound by an additional factor $(4 \pi)^2 \sim 10^2$.

\subsection{UV contribution to flavour violations}\label{sec:UVflavour}
It is equally important to consider the effect of four fermion interactions generated at the UV cutoff $\luv$ and to make sure that their presence does not reintroduce the flavour problem. We can rewrite the Lagrangian Eq.(\ref{eq:4 ferm UV}) responsible for the generation of light quark masses as
\begin{equation}\label{eq: 4 ferm CFT}
\mathcal{L}=\lambda^u(\luv)\ \bar{q}u\ \mathcal{O}+h.c.
\end{equation}
focusing on the up sector and neglecting flavour indices for brevity; quark masses are then given by
\begin{equation}
\mathcal{L}=\lambda^u(\Lambda_{HC})\Lambda_{HC}^{[\mathcal{O}]}\frac{v}{f}\,\bar qu+h.c.=4\pi\lambda^u(\Lambda_{HC})\Lambda_{HC}^{[\mathcal{O}]-1} v\,\bar q u+h.c.\,,
\end{equation}
employing $\langle\mathcal{O}\rangle=\Lambda_{HC}^{[\mathcal{O}]} v/f$ and $\Lambda_{HC}\simeq4\pi f$, $[\mathcal{O}]$ being the dimension of the operator $\mathcal{O}$. If we assume that the theory is an interacting CFT between $\luv$ and an infrared fixed point $\Lambda_{HC}$, where $\SO(5)$ is broken to $\SO(4)$, $[\mathcal{O}]$ is nearly scale independent and the running of $\lambda^u$ is well captured by
\begin{equation}
\lambda^u(\Lambda_{HC})=\lambda^u(\luv){\left(\frac{\Lambda_{HC}}{\luv}\right)}^{[\mathcal{O}]-1}\,.
\end{equation}
Moreover we can define a dimensionless coupling $\bar{\lambda}^u(\luv)=\lambda^u(\luv)\luv^{[\mathcal{O}]-1}$. Putting everything together we find quark masses
\begin{equation}\label{eq:masses CFT}
4\pi\bar\lambda^u(\luv){\left(\frac{\Lambda_{HC}}{\luv}\right)}^{2([\mathcal{O}]-1)}v\,.
\end{equation}
Requiring that Eq.(\ref{eq:masses CFT}) reproduces the charm mass, or equivalently the charm Yukawa times $v$, and imposing $\bar\lambda^u(\luv)\leq4\pi$, namely perturbativity at the scale where the operator $\mathcal{O}$ is generated, we have
\begin{equation}
{\left(\frac{\Lambda_{HC}}{\luv}\right)}\geq{\left(\frac{y_c}{16\pi^2}\right)}^\frac{1}{{2([\mathcal{O}]-1)}}\simeq6\times10^{-5}
\end{equation}
choosing $[\mathcal{O}]=1.5$ \cite{Rattazzi:2008pe} (see also \cite{Caracciolo:2014cxa} and references therein). Since $\Lambda_{HC}\simeq4\pi f\simeq10$ TeV we get $\luv\lesssim10^5$ TeV. Therefore four fermion interactions of the form
\begin{equation}
\mathcal{L}=\frac{1}{\luv^2}(\bar q q)^2+h.c.
\end{equation}
do not reintroduce any flavour problem as large enough suppression scales are allowed. 
These four-fermion interactions are a generic prediction of the physics responsible of Eq.(\ref{eq: 4 ferm CFT}) and from an effective theory point of view they can be suppressed only decoupling the UV cutoff: we avoid tensions typical of technicolour theories because we need to fix the charm -- and not the top -- mass. Finally notice that the same line of reasoning is applied to the down sector and a single cutoff $\luv$ is consistent since $m_c\sim m_b$.

\subsection{Summary and discussion}

In the up sector, top partial compositeness and additional Yukawa interactions can be combined safely from the point of view of flavour observables.
This strongly relies on hierarchies in the mass matrices which are generated by the two different mass sources. 
For what concerns other tests, corrections to EW precision parameters $S$ and $T$ can be computed as in composite Higgs models, with the additional Yukawa interaction playing a very minor role. We then refer to the literature for estimates, such as \cite{Giudice:2007fh,Barbieri:2007bh,Contino:2010rs} and \cite{Grojean:2013qca,Matsedonskyi:2014iha}. We content ourselves noting that, generically, with our choice of $f$ and $\Lambda_{HC}$, EW tests can be satisfied.

Since a couple of years, the Higgs boson is a new player in constraining new models via the knowledge we have of its couplings.
In composite Higgs models, relative deviations in its couplings to quarks are given by non linearities and, henceforth, they depend on the form of the interactions with the strong sector: in the case at hand, Eq.(\ref{eq: 4 ferm Yukawa}), the correction is universal and it has the form
\begin{equation}
\frac{y_{SM}-y}{m/v}\simeq1-\frac{1-2 s_\epsilon^2}{\sqrt{1-s_\epsilon^2}}\simeq0.15\,.
\end{equation}
This value is still allowed for the $h\bar{b}b$ coupling \cite{Giardino:2013bma}. For light quarks the Yukawa couplings are not constrained with the same precision.

The only sector where anarchic UV mass terms are in tension with data is the down sector: here, flavour bounds require the mixing angles to be small, so that a certain amount of alignment seems to be necessary.
A combined analysis of all the results we collected is in order:
{\small \begin{eqnarray}\label{eq: bound collection}
\mbox{Z boson FCNCs, Eq.(\ref{eq: down Z2})}&\Rightarrow & 
|V_{dL33}^\ast V_{dL13}|<10^{-1}\,,\quad|V_{dL33}^\ast V_{dL23}| <10^{-1/2}\,,\quad |V_{dL13}^*V_{dL23}|<10^{-5/2}\,,\nonumber\\
\mbox{CKM unitarity, Eq.(\ref{eq: CKM uni})}&\Rightarrow& 
{|V_{dL13}|}<10^{-1}\,,\quad{|V_{dL23}|}<10^{-1/2}\,, \nonumber \\
\mbox{Scalar resonance, Eq.(\ref{eq: phi down z bound})}&\Rightarrow&
|z_4^{db}|<1\div 10^{-2}\,,\quad|z_4^{sb}|< 1 \div 10^{-1/2}\,,\quad |z_4^{ds}|<10^{-4} \div 10^{-6}\,,\\
\mbox{Vector resonance, Eq.(\ref{eq: heavy vector})}&\Rightarrow&
|V_{dL33}^\ast V_{dL31}|<10^{-1} \div 10^{-3}\,,\quad |V_{dL33}^\ast V_{dL32}|<1 \div 10^{-2}\,, \nonumber \\ 
&  & \quad |V_{dL32}^\ast V_{dL31}|<10^{-3} \div 10^{-5}\,.\nonumber
\end{eqnarray}}
The range in the case of resonances is due to the unknown value of the masses and couplings of the resonances.
The only constraints that derive directly from partial compositeness in the up-sector are the ones from CKM unitarity: however, they require a quite mild hierarchy in the down-sector mixing matrix, especially in the first generation.
It should also be noted that the effect scales like $M_\ast^{-2}$, so increasing the mass of the top partners can help releasing the tension. The strongest constraints come from higher order operators (in the case of the Z boson FCNCs) and heavy resonances, thus their presence is more model dependent. Nevertheless, there is no way to avoid such contributions in general.

A possible simple way to contemporarily fulfill all the limits is to have $V_{dL13}=0$ and $|V_{dL23}|<10^{-2}$, with $V_{dL33}=O(1)$ and generic $V_{dR}$: we would not regard to this choice as particularly fine tuned; moreover many other possibilities are available. A very special case would be to have the down mass matrix hierarchical as it happens in the up sector, forcing the unitary transformations to have the form
\begin{equation}\label{eq: down hie}
V_{dL,R}\sim\left(\begin{array}{ccc}
O(1)&O(1)&O(\frac{m_s}{m_b})\\
O(1)&O(1)&O(\frac{m_s}{m_b})\\
O(\frac{m_s}{m_b})&O(\frac{m_s}{m_b})&1
\end{array}\right)\,.
\end{equation}
This in general is not completely satisfactory because the constraints on the coefficients $C_1^{bd,sd}$ of down-type operators coming from the exchange of heavy vector resonances generate a residual tension, as they may be one order of magnitude larger than the bounds; however an agreement with experiments can be obtained by varying the mass and couplings of the resonances.
The structure in Eq.(\ref{eq: down hie}) would be a consequence of $\lambda^d_{33}\gg\lambda^d_{\alpha\beta}$, such that $m_{D,33}\simeq m_b$ while $m_{D,\alpha\beta}\simeq m_s$ for all the other entries.  Notice that this would change all the coefficients discussed, for instance $\Sigma_d$ or the couplings in Eq.(\ref{eq: Phi down}): we checked that this would also satisfy all the experimental bounds. In the down sector mass terms all originate from the same operators and in principle no hierarchy is expected. In the following section we will show a possible way to generate such hierarchy by further extending the model and making the bottom partially composite, and fixing the other down masses to be of the order of the strange mass. This would make the down sector similar to the up sector, with a clear distinction between the $\{3,3\}$ entry and the others in the mass matrix and the form of the diagonalizing $V_{dL,dR}$ would be a consequence. We also point out that the simultaneous holding of Eq.(\ref{eq: diagon masses}) and Eq.(\ref{eq: down hie}) for $V_{uL}$ and $V_{dL}$ respectively is in agreement with the observed values of the third family entries of the CKM matrix.

It is instructive to revisit the limits collected in Eq.(\ref{eq: bound collection}) allowing the entries of $V_{dL}$ to be $O(1)$ complex numbers, apart from ${|V_{dL13}|}<10^{-1}$ and ${|V_{dL23}|}<10^{-1/2}$ because of CKM unitarity: this in turn implies $|V_{dL31,32}|\lesssim10^{-1/2}$ because of $V_{dL}$ unitarity. We report here the values of the masses of heavy resonances probed by reconsidering the processes discussed above under this viewpoint:

{\small \begin{align}
\mbox{Z boson FCNCs, Eq.(\ref{eq: down Z2})}&\Rightarrow  
m_V>(3\mbox{ TeV}){\left(10^{1/2}\right)}^{1/4}=4{\mbox{ TeV}}\,,\\
\mbox{Scalar resonance, Eq.(\ref{eq: phi down z bound})}&\Rightarrow
g_B^2\,{\left(\frac{1\mbox{ TeV}}{m_\Phi}\right)}^2{\left(\frac{1\mbox{ TeV}}{M_*}\right)}^2<10^{-5}\
\Rightarrow m_\Phi=M_*>\sqrt{g_B}\,17{\mbox{ TeV}}\,,\nonumber\\
\mbox{Vector resonance, Eq.(\ref{eq: heavy vector})}&\Rightarrow
g_*^2\,{\left(\frac{1\mbox{ TeV}}{M_*}\right)}^4{\left(\frac{1\mbox{ TeV}}{m_V}\right)}^2<10^{-8}\Rightarrow m_V=M_*>g_*^{1/3}21\mbox{ TeV}\,.\nonumber
\end{align}}

To conclude, we briefly address issues related to CP violation.
So far, we neglected all phases and treated all parameters as real: the suppressions we find are enough also for the imaginary parts.
However some flavour conserving CP violating processes such as the neutron EDM might be enhanced. The current experimental bound is \cite{Baker:2006ts}
\begin{equation}
|d_n|<2.9\times10^{-26} \,e\mbox{ cm}\quad (90\,\%\mbox{CL})\,.
\label{eq:dnbound}
\end{equation}
New physics effects can be sizable, indeed the neutron EDM receives contributions from the quarks EDMs.
The effects of partial compositeness have been investigated elsewhere in the literature \cite{Redi:2011zi,Agashe:2004cp,Konig:2014iqa}. 
We estimated the order of magnitude of the quarks EDMs, $d_{u,d}$, retaining only the fermions included in $\xi_{\uparrow}$ and $\xi_{\downarrow}$ and restricting to one loop diagrams with $Z$, $W$ and Higgs bosons.  Neglecting QCD running effects, expected to be $O(1)$, we find generic contributions to $d_{u,d}$ in the range of $10^{-21}\div10^{-24}\,e$ cm, thus up to five orders of magnitude above the experimental bound in Eq.(\ref{eq:dnbound}). Particular choices of parameters or unitary matrices $V_{dL,R}$ might improve the situation for the EDM of the down quark. In the up sector there are some fixed contributions, coming from Higgs exchange, and to properly account for them we have to assume that additional cancellations are at work or that the relative phase between $V_{uL31}^*$ and $V_{uR31}$ is small, less than $10^{-4}$. A full understanding of the neutron EDM constraint relies on a complete theory of flavour and on the knowledge of the strongly interacting sector, and therefore is outside the scope of our effective parametrization: for this reason we do not include it in our global analysis.

\section{Bottom mass}\label{sec: bottom semiPC}

We have focused so far on top partial compositeness and on direct Yukawa couplings. There is the possibility to propagate EWSB to the bottom quark if it linearly couples to composite operators as well, as a variant of what discussed above:
\begin{equation}
\mathcal{L}\supseteq \bar{q}_{3L} \mathcal{O}_{q_L} + \bar{b}_R \mathcal{O}_{b_R} + h.c.
\end{equation}
We choose the minimal option consisting in introducing linear mixings for both left-handed and right-handed fermions with the same composite resonance of the effective theory of the strong sector. This mechanism differs from the usually considered  partially composite bottom scenario in which additional composite  fermions are introduced as bottom partners.
In our proposal, the right-handed bottom develops a linear mixing with the same partner, $B$, which also mixes to the left-handed bottom, with the difference that the mixing in the right-handed sector vanishes when the EW symmetry is restored.

Indeed, given that the left-handed doublet $q_{3L}$ already mixes with the strong sector to give rise to the top mass, we do not need to add any new resonance. We can just complement Eq.(\ref{eq:Lag tot PC}) and Eq.(\ref{eq: 4 ferm Yukawa}) with the following effective operator, written in a formally $\SO(5)$ invariant way:
\begin{equation}\label{eq: b semi PC}
\mathcal{L}=y_R f \,\bar{\psi}_LU^td_{3R}^{14}\Sigma+h.c.=\frac{1}{2}y_R f s_\theta \bar{B}_Lb_R+h.c.\,,
\end{equation}
where the last equality holds in the unitary gauge, $\psi$ is the quark partner five-plet defined in Eq.(\ref{def psi=4+1}) that contains the bottom partner $B$, and $d_{3R}^{14}$ is a spurion formally transforming as the $\bf 14$ of $\SO(5)$, whose dynamical component is only the right-handed bottom:
\begin{equation}
d_{3R}^{14}=\frac{b_R}{2\sqrt{2}}\left(\begin{array}{ccccc}
0&0&1&i&0\\
0&0&-i&1&0\\
1&-i&0&0&0\\
i&1&0&0&0\\
0&0&0&0&0
\end{array}\right)\,.
\end{equation}
With this embedding, its U(1)$_X$ charge is $2/3$, matching the charge of $\psi$. An equivalent term could have been written embedding $b_R$ in a different representation, as the $\bf 10$ for instance, or in any other $\SO(5)$ representation whose decomposition to $\SO(4)$ contains the $\bf 4$
\footnote{The term in Eq.(\ref{eq: b semi PC}) can be formally rewritten as $\Tr[\bar{Q}_{14}d_{3R}^{14}]$, defining $Q_{14}=U(Q_1+Q_4+Q_9)U^t$, assuming that in the effective theory $Q_4$ mixes with the four-plet $Q$ defined in Eq.(\ref{def psi=4+1}) or directly identifying $Q$ and $Q_4$ and then decoupling the unnecessary components of $Q_{14}$, namely $Q_1$ and $Q_9$. This also suggests one way to UV complete this Lagrangian in the fundamental theory. We do not study this particular realization in detail but in the following section we explain how to generalize the results of our analysis to encompass cases like this.}
. Eq.(\ref{eq: b semi PC}) is the most general term that we could add, in particular we can always go to the basis where only one out of the three right-handed down-type quark, consequently defined as $b_R$, couples to $\psi_L$.

Because of the partial compositeness of $q_{3L}$ and taking into account the already present elementary Yukawas we obtain
\begin{equation}
\mathcal{L}=\bar{d}_{\alpha L}m^d_{\alpha\beta}d_{\beta R}+h.c.\,,\quad m^d=m^d_{UV}\frac{s_{2\epsilon}}{2}+\Pi \frac{fy_Rs_{\phi L}}{2}s_\epsilon\,.
\end{equation}
The second mass term in $m^d$ is generated by the bottom partial compositeness. We exploit here this mechanism to mimic what happened in the up-type sector: for this reason we require $f y_R s_{\phi L}s_\epsilon/2\simeq m_b$, fixing $y_R \sim m_b/f=O(10^{-3})$, and we let the elementary Yukawa to take into account the strange (and down) mass, $|m^d_{UV}|\sim m_s\ll m_b$. This hierarchy generates the structure presented in Eq.(\ref{eq: down hie}) for the down sector and it introduces two small quantities: $m_s/m_b$ and $m_b/f$, which alleviate the need for the alignment in down sector Yukawas.
Notice that this is a consequence of two different origins of the masses and it does not rely on the specific mechanism mediating EWSB to the bottom.
We have checked that this modification is safe from the point of view of the observables of Section \ref{sec:phenosec}, reconsidering the whole discussion including Eq.(\ref{eq: b semi PC}). Remarkably contributions to $Z\bar{b}b$ couplings are under control because we do not introduce additional bottom partners as well as we keep the custodial $Zb\bar{b}$ symmetry~\cite{Contino:2006qr} for left-handed coupling as before: at tree level
\begin{equation}
\delta g_{Zb_L}=0\,,\quad\delta g_{Zb_R}=-\frac{gs_{2\epsilon}^2}{8c_W}{\left(\frac{y_{L4}f m^d_{{\rm UV}33}+M_4y_Rf}{M_4^2+y_{4L}^2f^2}\right)}^2 \simeq-2\frac{g}{c_W}\frac{c_{\phi L}^4}{s_{\phi L}^2}{\left(\frac{m_{b}}{M_\ast}\right)}^2\,.
\end{equation}
A quantitative change is present in the bottom Yukawa: since the dominant contribution has a new spurionic structure we get
\begin{equation}
\frac{y_{SM}-y}{m_b/v}\simeq1-\sqrt{1-s_\epsilon^2}\simeq\frac{1}{2}s_\epsilon^2\simeq0.05
\end{equation}
for the deviation in $h\bar{b}b$ coupling. Note that the operators $Q_1^{d_\alpha d_\beta}$ are still induced with the same coefficients as in Eq.(\ref{eq:heavy vectors C}), and henceforth the coefficients $C_1^{bd,sd}$ suffer from the same $O(10)$ tension, which can be resolved by an extra suppression coming from the masses and couplings of the resonances.

\section{Generalisation of the results} \label{sec:general}

The results we presented in the previous sections apply to the minimal scenario, however the source of suppression of the flavour violating effects is quite generic. In this section, we show how the results can be generalised to cases with top partners in more complicated representations, and in the case of less minimal cosets.

\subsection{Additional top partners}\label{sec:AddTopP}

The discussion carried out so far can be generalised to other cases, where on top of mass terms of the form Eq.(\ref{eq: 4 ferm Yukawa}) we have linear mixing with different partners. Allowing for larger representations of top partners, we restrict to custodians~\cite{Agashe:2006at} for a zero tree level correction to $Z\bar{b}_Lb_L$: the minimal options are ${\bf 10}={\bf 4}+{\bf 6}$ and ${\bf 14}={\bf 1}+{\bf 4}+{\bf 9}$~\cite{DeSimone:2012fs}, decomposed in $\SO(4)$ irreducible representations. Both $\bf 6$ and $\bf 9$ contain partners with $Q=-1/3$ and the quantum numbers of $b_R$: they can, therefore, couple to $b_L$ and the physical Higgs. After going to the mass basis and integrating out heavy fields we are left with a structure similar to the minimal setup, but with additional flavour violating Higgs couplings. For definiteness we focus on partners in the $\bf 9$ with a mass $M_9$ and mixing $y_{L9}$\footnote{Top partners in the $\bf 14$ representation result in less fine tuned Higgs potential compared with the $\bf 5$ case \cite{Panico:2012uw}: $4$ and $5$ dimensional models featuring them can be found respectively in \cite{Pomarol:2012qf} and in \cite{Pappadopulo:2013vca}.}. In this extended set-up, in the down sector we obtain the following flavour-violating couplings of the Higgs:
\begin{eqnarray}
\mathcal{L}&\simeq&-\frac{s_{\phi9 L}^2c_{\phi L}^3}{c_{\phi9 L}^2}\frac{ \epsilon^2}{f}\ h\ \bar{b}_L \left( m^d_{{\rm UV}31} d_R+m^d_{{\rm UV}32}s_R+m^d_{{\rm UV}33} b_R\right)\\&\propto&
\left(
\begin{array}{ccc}
\bar{d}_L&\bar{s}_L&\bar{b}_L
\end{array}\right)
\small{\left(\begin{array}{ccc}
0&0&0\\
0&0&0\\
m^d_{{\rm UV}31}&m^d_{{\rm UV}32}&m^d_{{\rm UV}33}
\end{array}\right)}
\left(\begin{array}{c}
d_R\\
s_R\\
b_R
\end{array}\right)\,,\nonumber
\end{eqnarray}
where $s_{\phi9L} = y_{L9} f /\sqrt{M_9^2 + y_{L9}^2 f^2}$.
We neglect here the subdominant contributions, proportional to $m_b^3$, similar to $B_d$ reported for the minimal case discussed above. The third column of this matrix would be different from zero if we had couplings of the form $\bar{Q}_L b_R$ to start with: this does not happen as long as the right bottom is completely elementary. 

These flavour violating Higgs couplings affect meson observables through the following dimension six operators
\begin{eqnarray}\label{eq: HIggs FCNC 14}
\mathcal{L}&\simeq&\frac{1}{m_H^2}{\left(\frac{\epsilon m_b}{f}\right)}^2\left[z_4^{db}\mathcal{Q}_4^{db}+z_4^{sb}\mathcal{Q}_4^{sb}+z_4^{ds}\mathcal{Q}_4^{ds}\right]\simeq \frac{10^{-4}}{\mbox{TeV}^2} \left[z_4^{db}\mathcal{Q}_4^{db}+z_4^{sb}\mathcal{Q}_4^{sb}+z_4^{ds}\mathcal{Q}_4^{ds}\right]\,.
\end{eqnarray}
Experimental results imply then
\begin{equation}
|z_4^{db}|<10^{-3}\,,\quad |z_4^{sb}|<10^{-1}\,,\quad|z_4^{ds}|<10^{-7}\,.
\end{equation}
Therefore the simple assumptions $V_{dL13}=0$ and $|V_{dL23}|\lesssim10^{-1}$ would be enough, in analogy with the the minimal case of partners in the $\mathbf 5$.

If we assume that the quark mass matrix of the down sector is hierarchical, the $\{3,3\}$ entry of order $m_b$ being parametrically larger than the others of order $m_s$, and that therefore both $V_{dL,R}$ have the form Eq.(\ref{eq: down hie}), we have that Eq.(\ref{eq: HIggs FCNC 14}) reads
\begin{equation}
\frac{1}{m_h^2}{\left(\epsilon\frac{m_s}{f}\right)}^2\left[\mathcal{Q}_4^{bd}+{Q}_4^{bs}+{\left(\frac{m_s}{m_b}\right)}^2\mathcal{Q}_4^{ds}\right]\simeq\frac{10^{-7}}{\mbox{ TeV}^2}\left(\mathcal{Q}_4^{bd}+{Q}_4^{bs}+5\times10^{-4}\mathcal{Q}_4^{ds}\right)\,,
\end{equation}
thus passing all bounds. The same comment made at the end of Section \ref{subsec: FCNC MCHM5} applies here: a partially composite bottom quark would dynamically explain this hierarchy.

\subsection{Non minimal cosets and more general formalism}\label{sec: general argum}

Going beyond the minimal coset considered in the previous sections may seem a considerable complication.
There are many possible viable cosets (for an up to date review, see Ref.~\cite{Bellazzini:2014yua}), some of which can be obtained in simple dynamical UV completions with only fermionic components~\cite{Galloway:2010bp,Ferretti:2014qta}.
However, in models where a custodial symmetry is present, one always has the following situation:
\begin{equation}
\mathcal{G} \to \mathcal{H} \supset SO(4) \supset SU(2)_L \times U(1)_Y
\end{equation}
where $\mathcal{G}$ is the global symmetry broken to $\mathcal{H}$ by the dynamics, and SO(4) is the custodial symmetry.
Thus, while in principle one needs to consider top-partners in full representations of $\mathcal{H}$, the coupling to the top (and bottom) can be written in terms of subcomponents transforming under SO(4) alone.
It is now useful to analyse this scenario by use of the formalism developed to describe the interactions of general vector-like quarks, like in \cite{delAguila:2000aa,delAguila:2000rc} and \cite{Cacciapaglia:2011fx,Buchkremer:2013bha}, that mix to SM ones via couplings to the Higgs. The idea is that the couplings to light quarks can be characterised by the minimum number of ``Higgs doublets" needed to couple them to the SM elementary fields in a gauge invariant way. Even in the non-linear case we are interested in, this will be an index of the suppression in powers of $\epsilon$ of the given term.
For instance, a composite fermion that is a doublet of SU(2)$_L$ with hypercharge $Y=1/6$ can have a direct coupling at order $\epsilon^0$ with the elementary left-handed doublets, and coupling suppressed by at least one power of $\epsilon$ to the singlets.
A composite doublet with hypercharge $Y=7/6$ will have a coupling of order $\epsilon^2$ to the elementary doublets, and $\epsilon$ to the top singlet. In general, composite fermions that have semi-integer SU(2)$_L$ weak isospin (such as doublets and quadruplets), will have couplings to the left-handed elementary doublet which are even powers of $\epsilon$, and to the right-handed elementary singlets with are odd powers of $\epsilon$.
For composite fermions with integer weak isospin (such as singlets and triplets), the coupling is even in $\epsilon$ to the right-handed singlets and odd to left-handed doublets.
Note that non minimal cosets may contain more than one Higgs doublet, so $\epsilon$ should really be thought of as the spurion than breaks SU(2)$_L$, and the sharing of this among various doublets can be expressed in terms of the effective couplings.

As a first example, we can consider a situation containing the same composite fermions as in Section~\ref{sec: MCHM5 plus TC}, {\emph{i.e.}} a doublet $Q = \{T,B\}$ with hypercharge $1/6$, a doublet $X = \{X_{5/3}, X_{2/3}\}$ with hypercharge $7/6$ and a singlet $\tilde{T}$ with hypercharge $2/3$.
Following this argument, the general mass matrix for the up sector can be written as
\begin{equation}\label{eq:up mass G}
M_{\rm up} = \left(
\begin{array}{cccccc}
 &  &  & 0 & 0 & 0 \\
 & m_{\rm UV}^{\alpha\beta}\, \epsilon  & & 0 & 0 & 0 \\ 
 & &  & Y_{LQ} & Y_{LX}\, \epsilon^2 & - Y_{L\tilde{T}}\, \epsilon \\
0 & 0 & Y_{RQ}\, \epsilon & M_4 & 0 & 0 \\
0 & 0 & - Y_{RX}\, \epsilon & 0 & M_4 & 0 \\
0 & 0 & Y_{R\tilde{T}}  & 0 & 0 & M_1
\end{array} \right)
\end{equation}
in the interaction basis $\xi_{\uparrow}=\begin{pmatrix}
u&c&t&T&X_{2/3}&\tilde{T}
\end{pmatrix}$,
where $T$ and $X_{2/3}$ belong to a SO(4) bi-doublet, and have therefore the same mass $M_4$.
The case we discussed in the previous section, {\emph{i.e.}} the mass matrix $M_{\rm up}$ obtained from Eq.(\ref{eq:Lag tot PC}), can be seen as a particular case, with $Y_{LQ} = f y_{L4} \cos^2 \epsilon/2$, $Y_{LX} = f y_{L4} \frac{\sin^2 \epsilon/2}{\epsilon^2}$, etc. The results we present here, therefore, are another way to look at the analysis we performed in Section~\ref{sec: MCHM5 plus TC}.
In the down sector, in the interaction basis $\xi_{\downarrow}=\begin{pmatrix}
d&s&b&B
\end{pmatrix}$, we can write:
\begin{equation}\label{eq:down mass G}
M_{\rm down} = \left(
\begin{array}{cccccc}
 &  &  & 0 \\
 & m_{UV}^{\alpha\beta}\, \epsilon  & & 0  \\ 
 & &  & Y_{LQ} \\
0 & 0 & Y_{RQ,b}\, \epsilon & M_4  
\end{array} \right)\,.
\end{equation}
Note that, in the minimal \SO(5)/\SO(4) model, $Y_{RQ,b}$ is usually absent because the elementary $b_R$ is embedded into a 5-plet of SO(5) whose U(1)$_X$ charge does not match with the charge of the composite 5-plet. As this term is allowed by the SM symmetries, it should always be possible to write an operator that generates it: in the minimal case, this requires embedding $b_R$ into a spurion transforming like a 2-index representation of SO(5), as we showed in Section~\ref{sec: bottom semiPC}.

Here we have explicitly factored out the minimal power of $\epsilon$ necessary to write a gauge invariant term, so that all the objects like $m_{\rm UV}$ and the $Y$'s are functions of $\epsilon^2$ starting with a constant in a small-$\epsilon$ expansion.
The masses $M_4$ and $M_1$ are vector-like masses of the two multiplets: depending on the coset, they may be equal if the two fermions belong to the same multiplet of $\mathcal{H}$.
The mass matrix in the up sector can be, in general, diagonalised by two independent 6$\times$6 matrices:
\begin{equation}
U_{L,u} \cdot M_{\rm up} \cdot U_{R,u}^\dagger = M_{\rm up}^{diag}\,.
\end{equation}
Nevertheless, not all entries of $U_{L,u}$ and $U_{R,u}$ are relevant for the flavour physics.

In the interaction basis, the couplings of the $Z$ boson to the up sector are given by Eq.\eqref{eq: Z up MCHM}.
Setting for the time being the couplings $c_{L,R}$ to zero, once we pass to mass eigenstate basis we get
\begin{eqnarray}
g_{ZL}^{mass} = U_{L,u} \cdot g_{ZL} \cdot U_{L,u}^\dagger &=& g_{ZL}^{\rm SM} \delta^{i,j} - \frac{g}{\cos \theta_W} \left( U_{L,u}^{i5} U_{L,u}^{\ast,j5} + \frac{1}{2} U_{L,u}^{i6} U_{L,u}^{\ast,j6}  \right)\,, \\
g_{ZR}^{mass} = U_{R,u} \cdot g_{ZR} \cdot U^\dagger_{R,u} &=& g_{ZR}^{\rm SM} \delta^{i,j} + \frac{g}{\cos \theta_W} \left(\frac{1}{2} U_{R,u}^{i4} U_{R,u}^{\ast,j4} - \frac{1}{2} U_{R,u}^{i5} U_{R,u}^{\ast,j5} \right) \,;
\end{eqnarray}
where $i,j = 1\dots 6$, and $g_{ZL/R}^{\rm SM}$ are the appropriate SM couplings.
Note that the form of this couplings is very similar to a scenario with generic vector-like fermions (see the Appendix of Ref.~\cite{Buchkremer:2013bha}). 
In the SM sector, $g_{Z}^{\alpha \beta}$ with $\alpha,\beta = 1,2,3$, the deviations in the $Z$ couplings are proportional to the matrix elements $U_{L,u}^{\alpha5/6}$ and $U_{R,u}^{\beta4/5}$.
It can be shown that the leading contribution to such terms in the mixing matrices is
\begin{eqnarray}
& U_{L,u}^{\alpha5} \sim \left( \frac{m_c}{m_t} \epsilon^2, \ \frac{m_c}{m_t} \epsilon^2, \ \epsilon^2 \right)\ \qquad U_{L,u}^{\alpha6} \sim \left( \frac{m_c}{m_t} \epsilon, \ \frac{m_c}{m_t} \epsilon, \ \epsilon \right)\ & \\
& U_{R,u}^{\alpha4} \sim \left( \frac{m_c}{m_t} \epsilon, \ \frac{m_c}{m_t} \epsilon, \ \epsilon \right)\ \qquad U_{R,u}^{\alpha5} \sim \left( \frac{m_c}{m_t} \epsilon, \ \frac{m_c}{m_t} \epsilon, \ \epsilon \right)\ &
\end{eqnarray}
where $\epsilon = v/f \sim m_t/M_\ast$, $M_\ast$ being a generic top partner mass.
These features are very generic: the $\epsilon$ factor comes from the fact that the mixing is due to the EWSB, while the $m_c/m_t$ factors come from the hierarchy in the SM quark mass matrix between the top mass induced by partial compositeness and the $m_{\rm UV}$ contributions which are naturally of order $m_c$.
This structure generates a suppression of order $\epsilon^2 m_c^2/m_t^2 \sim 10^{-5}$ to deviation in the first two generations, and $\epsilon^2 m_c/m_t \sim 10^{-3}$ for flavour-violating top couplings.
Additional contributions arise from the composite nature of the new fermions, represented by the $c_{L,R}$ terms in Eq.\eqref{eq: Z up MCHM}. Those couplings arise from couplings of the pions and gauge bosons which connect two composite states belonging to different representations of the unbroken group $\mathcal{H}$. In the limit where the EW symmetry is unbroken, such terms necessarily vanish due to gauge invariance, therefore they ought to be proportional to at least one power of $\epsilon$. In our case, these terms connect the singlet $\tilde{T}$ to doublets, therefore they arise at order $\epsilon$, and have the form
\begin{equation}
\Delta g_{ZL/R} (c) = C_{L/R}^\ast \epsilon\ (\delta^{i4} \delta^{j6} + \delta^{i5} \delta^{j6}) + C_{L/R} \epsilon\ \ (\delta^{i6} \delta^{j4} + \delta^{i6} \delta^{j5})
\end{equation}
where we imposed the custodial symmetry, {\emph{i.e.}} same couplings for the two doublets, and $C_{L/R}$ are functions of $\epsilon^2$ starting with a constant. The above structure is the most general one, and it agrees with Eq.\eqref{eq: Z up MCHM} with $C_{L/R} = - \frac{c_{L/R} e}{\sqrt{2} c_W s_W} \frac{\sin \epsilon}{\epsilon}$.
In the mass eigenstate basis, they will generate an additional correction to the $Z$ couplings of the form
\begin{equation}
\Delta g_{ZL/R}^{mass} (c) = C_{L/R} \epsilon (U_{L/R,u}^{i6} U_{L/R,u}^{\ast,j4} + U_{L/R,u}^{i6} U_{L/R,u}^{\ast,j5} ) + C_{L/R,u}^\ast \epsilon (U_{L/R,u}^{i4} U_{L/R,u}^{\ast,j6} + U_{L/R,u}^{i5} U_{L/R,u}^{\ast,j6})\,,
\end{equation}
thus in the SM quark sector, the FCNCs are still proportional to the same mixing matrix elements, including
\begin{equation}
 U_{L,u}^{\alpha 4} \sim U_{R,u}^{\alpha 6} \sim \left( \frac{m_c}{m_t} , \ \frac{m_c}{m_t} , \ 1 \right)\,.
\end{equation}
These mixing angles are not suppressed by any power of $\epsilon$ as they involve composite fermions with the same quantum numbers as the elementary quarks, however the missing factor of $\epsilon$ is compensated by the coupling.
Therefore, we can say that the contribution of the $c_{L/R}$ terms always arises at the same level as the other effects, and are safely suppressed in the up sector.

In the down sector, the mass matrix can be diagonalised by 4$\times$4 unitary matrices
\begin{equation}
U_{L,d} \cdot M_{\rm down} \cdot U_{R,d}^\dagger = M_{\rm down}^{diag}\,.
\end{equation}
The bottom quark will now receive a contribution from the linear coupling to the composite partner
\begin{equation}
m_b \sim \frac{Y_{LQ} Y_{RQ,b} \epsilon}{\sqrt{M_4^2+Y_{LQ}^2}} + O(\epsilon^3)\,;
\end{equation}
if we want this to reproduce most of the bottom mass, then we need $Y_{RQ,b} \sim m_{b}$.
As the composite $B$ has the same quantum numbers as the elementary left-handed bottom, no corrections to the left-handed $Z$ couplings will arise, while in the right-handed sector
\begin{equation}
g_{ZR,d}^{mass} = U_{R,d} \cdot g_{ZR,d} \cdot U^\dagger_{R,d} = g_{ZR,d}^{\rm SM} \delta^{i,j} - \frac{g}{\cos \theta_W} \frac{1}{2} U_{R,d}^{i4} U_{R,d}^{\ast,j4}\,.
\end{equation}
For the mixing matrices, analogously to the up case, we obtain
\begin{equation}
U_{L,d}^{\alpha 4} = - s_{\phi_L} \left(\frac{m_{UV,d} \epsilon}{m_b}, \frac{m_{UV,d} \epsilon}{m_b}, 1\right)\,, \qquad U_{R,d}^{\alpha 4} = - \frac{m_b \cot \phi_L}{\sqrt{M_4^2+Y_{LQ}^2}} \left(\frac{m_{UV,d} \epsilon}{m_b}, \frac{m_{UV,d} \epsilon}{m_b}, 1\right)\,.
\end{equation}
We can see from these formulas that the flavour violation in the right-handed sector due to mixing to the composite states is suppressed by an extra factor $m_b/m_t \sim m_b/v$, while an hierarchy in the coupling to light generations can be naturally obtained if $m_{UV,d}\epsilon \sim m_s \ll m_b$.

Similar suppression factors also arise in the Higgs couplings.
This structure is enough to avoid bounds from flavour conserving and violating effects in the light generation, and in the top quark.

The charged current sector depends on the bottom sector as well.
In the minimal scenario under consideration, where only the left-handed bottom is partially composite, there is a single bottom partner belonging to a doublet, therefore the mass matrix can be diagonalised by two 4$\times$4 rotations $U_{L,d}$ and $U_{R,d}$.
The couplings of the $W$ are given in Eq.\eqref{eq: W MCHM}.
From it we can extract the 3$\times$3 SM CKM matrix
\begin{equation}
V_{CKM}^{\alpha\beta} = (v_L \cdot v_{L,d}^\dagger)^{\alpha\beta} + k \epsilon^2 v_L^{\alpha3} v_{L,d}^{\ast, \beta3}
\end{equation}
where $v_L$ and $v_{L,d}$ are unitary 3$\times$3 matrices sensitive to the diagonalisation of the masses in the SM sectors (similar to the rotation matrices in Eq.\eqref{eq:V for masses}), and $k$ is an order 1 coefficient.
In the up sector, the hierarchy between the partial composite top and light quarks generates a hierarchy on $v_L$ in a similar fashion as the one in Eq.\eqref{eq: diagon masses}.
The non-unitarity in the up sector is given, at leading order, by
\begin{equation}
(V_{CKM} \cdot V_{CKM}^\dagger )^{\alpha\beta} = \delta^{\alpha\beta} + 2 k \epsilon^2 v_L^{\alpha3} v_{L}^{\ast, \beta3} \
\end{equation}
thus the hierarchy in $v_L$ suppresses the effect in the light sector by a factor $\epsilon^2 m_c^2/m_t^2 \sim 10^{-5}$, which is well below the accuracy to which the unitarity is measured in the first and second generation cases~\cite{Agashe:2014kda} which is of order 0.5\%.
In the third component, one may expect effects of the order $\epsilon^2 \sim 0.1$, which are however compatible with the poorer direct determination of $|V_{tb}|=1.021\pm0.032$ \cite{Agashe:2014kda}.

In the down-sector, we similarly have
\begin{equation}
(V_{CKM}^\dagger \cdot V_{CKM} )^{\alpha\beta} = \delta^{\alpha\beta} + 2 k \epsilon^2  v_{L,d}^{\alpha3} v_{L,d}^{\ast, \beta3}\,.
\end{equation}
If the right-handed bottom is fully elementary, the $\epsilon^2 \sim 0.1$ suppression is not enough to limit unitarity violation in the light quark sector enough to pass the precise determination, which fares at 0.5\% level for the down and 5\% for the strange~\cite{Agashe:2014kda}.
One is therefore forced to consider a certain hierarchy in the $m_{UV}$  down masses.
Another possibility is to make the bottom partially composite, so that a hierarchy of order $m_s/m_b \sim 0.02$ is introduced: this would be enough to avoid bounds from CKM unitarity without introducing hierarchies in the UV sector generating elementary Yukawas.

One minimal possibility is to use the coupling $Y_{RQ,b}\, \epsilon$ in Eq.(\ref{eq:down mass G}) to give a contribution to the bottom mass from partial compositeness, so that naturally $m_{UV} \epsilon \sim m_s$.
Another possibility is to enlarge the down sector by adding partners with the correct quantum numbers to mix with $b_R$.
In both cases, a similar discussion of this matrix on the same footing as Eq.\eqref{eq:up mass G} is possible and should give similar results as in the top sector.
The main difference is that small mixings are needed in the partial composite bottom to reproduce the correct bottom mass (schematically, $Y_{RQ,b} \sim m_b \ll Y_{LQ} \sim m_t$), so that partial compositeness effects in the down sector should be suppressed  by a factor $m_b/M_\ast$.

The above discussion can be extended to any number and representation of composite fermions, as long as the spurion breaking SU(2)$_L$ comes from one (or more) doublets. Independently of the properties of the composite fermion, the mixing elements connecting it to the SM fermions have the form:
\begin{equation}
U_{L,u}^{\alpha i} \sim \left( \frac{m_c}{m_t} , \ \frac{m_c}{m_t} , \ 1 \right) \epsilon^{n_L}\,, \qquad U_{R,u}^{\alpha i} \sim \left( \frac{m_c}{m_t} , \ \frac{m_c}{m_t} , \ 1 \right) \epsilon^{n_R}\,.
\end{equation}
Depending on the SM quantum numbers, the following cases can occur:
\begin{itemize}
\item the composite fermion has the same quantum numbers as the elementary $t_R$: $n_L = 1$ and $n_R = 0$, thus the corrections to the left-handed $Z$ couplings arise at order $\epsilon^2$ with no corrections to the right-handed ones;
\item the composite fermion has the same quantum numbers as the elementary $t_L$: $n_L = 0$ and $n_R = 1$, thus the corrections to the right-handed $Z$ couplings arise at order $\epsilon^2$ with no corrections to the left-handed ones;
\item the composite fermion has semi-integer isospin: $n_L \geq 2$ and $n_R \geq 1$, thus the left-handed coupling is suppressed by at least $\epsilon^4$ while the right-handed one by $\epsilon^2$;
\item the composite fermion has integer isospin: $n_L \geq 1$ and $n_R \geq 2$, thus the left-handed coupling is suppressed by at least $\epsilon^2$ while the right-handed one by $\epsilon^4$.
\end{itemize}
Thus, corrections to the couplings of the $Z$ are always suppressed by at least two powers of $\epsilon$, enough, together with the $m_c/m_t$ factors, to avoid flavour bounds.
For the $c_{L/R}$ terms:
\begin{itemize}
\item if $c_{L/R}$ connects a composite state with integer isospin to one with semi-integer isospin, the coupling is suppressed by at least a factor $\epsilon$, while at least another factor of $\epsilon$ comes from the mixing matrices;
\item if $c_{L/R}$ connects composite states with same isospin, the coupling is suppressed by at least $\epsilon^2$, while the mixing matrices may carry no $\epsilon$ suppression.
\end{itemize}
In all cases, therefore, at least a factor $\epsilon^2$ appears, together with the $m_c/m_t$ factors from the flavour mixing. Thus, the $c_{L/R}$ couplings always appear at the same order as the other corrections.

An important caveat to this analysis is the presence of additional pNGBs in non-minimal cosets: these states may have linear couplings to the fermions generated by the same terms giving mass to them. Therefore, in the mass eigenstate basis, additional FCNCs may be generated. However, the details of these extra contributions depend on the details of the model, the coset and the form of the mass terms, both from partial compositeness and the UV contributions. Thus, their effects should be checked case by case.

\section{Conclusions}\label{sec:conclusion}

We have explored the possibility that the top quark mass has a different origin than the masses of the other quarks (and leptons). This scenario can naturally arise in models of pNGB Higgs, deriving for instance from a strongly interacting underlying dynamics.
In this scenario, partial compositeness for top quark is responsible for generating top quark mass and the Higgs potential.
An unrelated source of mass is represented by deformations of the strong sector that generate bilinear couplings suppressed by a scale heavier than the condensation scale.
In order to avoid four-fermion contributions to FCNCs generated at this high scale, one needs it to be above $10^{5}$ TeV,
which is enough to generate the bottom, charm and tau masses, but not the top mass, if the dynamics is near-conformal down to the condensation scale.
In our scenario, therefore, the contribution of the top compositeness is crucial to achieve a large enough top mass.

We showed that this scenario is compatible with bounds from precision measurements of the quark couplings and from flavour constraints, without the need to assume a flavour symmetry in the underlying dynamics for the up sector.
Hence, while the top is naturally singled out as the heaviest of the SM quarks, the direct Yukawa couplings can be anarchic.
This property is due to suppressions in the corrections of order $v^2/f^2$ in the top sector, and  ${(m_cv)}^2/{(m_tf)}^2$ in the light quark sector. The situation is different for the down quarks: in the case where partial compositeness
is not employed for
the bottom mass, we observe only corrections of order $v^2/f^2$, notably in the unitarity of the CKM matrix.
Therefore, in order to satisfy the bounds, one is forced to ask for a certain hierarchy (requiring some kind of alignment) in the mixing in the down sector, which is fully generated by direct couplings.
However, allowing the bottom to be partially composite, {\sl i.e.} adding a further source for the bottom mass, can ease the tension if the contribution of direct couplings is smaller than the bottom mass, for instance the strange mass, and a fully anarchic scenario becomes plausible. In this case, the hierarchy in the mixing in the SM fermion sector is enough to suppress effects in the unitarity of the CKM matrix and anarchic direct couplings are fully allowed. We explicitly implemented a very economical bottom partial compositeness, without any additional fermionic resonance beyond the top partners: in this novel scenario, the right-handed bottom mixes via the EWSB to the same composite partner that mixes with the left-handed elementary bottom. 
For what concerns other observables, we do not expect significant deviations in flavour processes and the estimated sensitivities for the next run of LHC are still above the expected top flavour violating decays rates. On the other hand the neutron EDM might need a dedicated study and could be a significant test bench.
It would be also interesting to address the generation of the flavour hierarchies of the SM, of the CKM and of the SM CP violating phase, points that cannot be completely decoupled and solved at high energy because they receive inputs from physics at the EW scale.

A smoking gun for our model will be the prediction of the presence of heavy coloured fermions, which will be further investigated in the collider experiments. A discovery of the vector-like top (and bottom) partners, along with the absence of light fermion partners at the LHC will be a genuine prediction of our model framework.

Finally, we want to emphasize that, while we analysed in detail the minimal case of the coset \SO(5)/\SO(4) with top partners belonging to a four-plet and singlet of \SO(4) (aka MCHM$_5$),
our results are quite general. We showed how they can be extended to cases where the top (and bottom) partners belong to larger representations of \SO(4) and cases where the coset is larger. Our conclusions are therefore rather solid under variations of the models.

\section*{Acknowledgements}

We thank Aldo Deandrea for precious discussions and for commenting on the manuscript.
This work was supported by the National Research Foundation of Korea (NRF) grant funded by the Korea government (MEST) (No. 2012R1A2A2A01045722) and also supported by Basic Science Research Program through the National Research Foundation of Korea (NRF) funded by the ministry of Education, Science and Technology (No. 2013R1A1A1062597). H.S. acknowledges the Portuguese FCT project PTDC/FIS-NUC/0548/2012. A.P. acknowledges IBS Korea under system code IBS-R017-D1-2015-a00. Finally, G.C. and H.C. acknowledge partial support from the Labex-LIO (Lyon Institute of Origins) under grant ANR-10-LABX-66, and support from the France-Korea Particle Physics Lab (FKPPL).

\begin{appendix}

\section{Model details and perturbative expansion}\label{secdetails}
\subsection{Model details}
In this appendix we present some more details on the model discussed in Section \ref{sec: MCHM5 plus TC}. The 3rd family part of this model (including the mass matrix, its diagonalization, and charged and neutral currents in the mass eigenbasis) has been discussed in Ref.~\cite{Backovic:2014uma} which we extend, here, by adding the light quark flavours.
The up sector mass matrix, in the field basis in which the Lagrangian Eq.(\ref{eq:Lag tot PC}) is written, is
\begin{equation}\label{eq: Mup appendix}
\footnotesize{
M_{\rm up}=
\begin{pmatrix}
\tilde{m}[\epsilon]_{11}&\tilde{m}[\epsilon]_{12}&\tilde{m}[\epsilon]_{13}&0&0&0\\
\tilde{m}[\epsilon]_{21}&\tilde{m}[\epsilon]_{22}&\tilde{m}[\epsilon]_{23}&0&0&0\\
\tilde{m}[\epsilon]_{31}&\tilde{m}[\epsilon]_{32}&\tilde{m}[\epsilon]_{33}&fy_{L4}\cos^2\dfrac{\epsilon}{2}&fy_{L4}\sin^2\dfrac{\epsilon}{2}&-f\dfrac{y_{L1}}{\sqrt{2}}\sin\epsilon\\
0&0&f\dfrac{y^\ast_{R4}}{\sqrt{2}}\sin\epsilon&M_4&0&0\\
0&0&-f\dfrac{y^\ast_{R4}}{\sqrt{2}}\sin\epsilon&0&M_4&0\\
0&0&fy_{R1}^\ast\cos\epsilon&0&0&M_1
\end{pmatrix}
} \, ,
\end{equation}
with $\tilde{m}[\epsilon]_{\alpha\beta}\equiv \dfrac{s_{2\epsilon}}{2}m^u_{{\rm UV}\alpha\beta}$ and $s_{2\epsilon}=\sin2\epsilon$. The up-type Yukawa couplings have two distinct contributions. One arises form the mixing Lagrangian $\mathcal{L}_{mix}$ given in Eq.(\ref{eq:Lag tot PC}) and can be extracted by differentiating $\mathcal{L}_{mix}$ with respect to $h$ and then setting $h=0$, which yields
\begin{equation}
\footnotesize{
Y_{\rm up}^{mix}=
\begin{pmatrix}
\tilde{y}[\epsilon]_{11}&\tilde{y}[\epsilon]_{12}&\tilde{y}[\epsilon]_{13}&0&0&0\\
\tilde{y}[\epsilon]_{21}&\tilde{y}[\epsilon]_{22}&\tilde{y}[\epsilon]_{23}&0&0&0\\
\tilde{y}[\epsilon]_{31}&\tilde{y}[\epsilon]_{32}&\tilde{y}[\epsilon]_{33}&-\dfrac{y_{L4}}{2}\sin\epsilon&\dfrac{y_{L4}}{2}\sin\epsilon&-\dfrac{y_{L1}}{\sqrt{2}}\cos\epsilon\\
0&0&\dfrac{y^\ast_{R4}}{\sqrt{2}}\cos\epsilon&0&0&0\\
0&0&-\dfrac{y^\ast_{R4}}{\sqrt{2}}\cos\epsilon&0&0&0\\
0&0&-y_{R1}^\ast\sin\epsilon&0&0&0
\end{pmatrix}
} \, ,
\end{equation}
where $\tilde{y}[\epsilon]_{\alpha\beta}\equiv c_{2\epsilon}\dfrac{m^u_{{\rm UV}\alpha\beta}}{f}$. The second contribution comes from the $d_\mu$-term of $\mathcal{L}_{comp}$ and the fact that $d_\mu^4\propto \partial_\mu h$. Integrating by parts and using the equations of motion we obtain
\begin{equation}
\footnotesize{
Y_{\rm up}^{comp}=
\begin{pmatrix}
0&0&0&0&0&0\\
0&0&0&0&0&0\\
0&0&0&\dfrac{c_R^\ast y_{L1}}{\sqrt{2}}\sin\epsilon&-\dfrac{c_R^\ast y_{L1}}{\sqrt{2}}\sin\epsilon&c_Ry_{L4}\cos\epsilon\\
0&0&-c_Ly^\ast_{R1}\cos\epsilon&0&0&-\dfrac{c_L M_1-c_R M_4}{f}\\
0&0&c_Ly^\ast_{R1}\cos\epsilon&0&0&-\dfrac{c_R M_4-c_LM_1}{f}\\
0&0&\sqrt{2}c_L^\ast y^\ast_{R4}\sin\epsilon&-\dfrac{-c_L^\ast M_4+c_R^\ast M_1}{f}&-\dfrac{c_L^\ast M_4-c_R^\ast M_1}{f}&0
\end{pmatrix}\,.
}
\end{equation}
For the down sector we have
\begin{equation}\label{eq: Mdown appendix}
\footnotesize{
M_{\rm down}=
\begin{pmatrix}
\tilde{m}[\epsilon]_{11}&\tilde{m}[\epsilon]_{12}&\tilde{m}[\epsilon]_{13}&0\\
\tilde{m}[\epsilon]_{21}&\tilde{m}[\epsilon]_{22}&\tilde{m}[\epsilon]_{23}&0\\
\tilde{m}[\epsilon]_{31}&\tilde{m}[\epsilon]_{32}&\tilde{m}[\epsilon]_{33}&fy_{L4}\\
0&0&0&M_4
\end{pmatrix}\,,\quad
Y_{\rm down}=
\begin{pmatrix}
\tilde{y}[\epsilon]_{11}&\tilde{y}[\epsilon]_{12}&\tilde{y}[\epsilon]_{13}&0\\
\tilde{y}[\epsilon]_{21}&\tilde{y}[\epsilon]_{22}&\tilde{y}[\epsilon]_{23}&0\\
\tilde{y}[\epsilon]_{31}&\tilde{y}[\epsilon]_{32}&\tilde{y}[\epsilon]_{33}&0\\
0&0&0&0
\end{pmatrix}\,.
}
\end{equation}
The neutral currents mediated by the $Z$ boson yield the interaction Lagrangian
\begin{align}
\begin{split}
\mathcal{L}\supset
Z_\mu 
\overline{
\xi}_{\uparrow L,R}\gamma^\mu
A_{NC}^{tL,R}
\xi_{\uparrow L,R}+Z_\mu 
\overline{
\xi}_{\downarrow L,R}\gamma^\mu
A_{NC}^{bL,R}
\xi_{\downarrow L,R}\, ,
\end{split}
\end{align}
with $\xi_{\uparrow,\downarrow}$ defined in Eq.(\ref{eq:xidef}), and with the associated flavour matrices

{\footnotesize{
\begin{align}\label{eq: Z up MCHM}
\begin{split}
A_{NC}^{tL,R}=&
\begin{pmatrix}
\dfrac{e}{c_Ws_W}\left(\dfrac{\delta^L}{2}-\dfrac{2s_W^2}{3}\right)\mathbb{I}_3& \mathbb{O}_3^T & \mathbb{O}_3^T  & \mathbb{O}_3^T  \\
\mathbb{O}_3&\dfrac{e}{c_Ws_W}\left(\dfrac{1}{2}-\dfrac{2s_w^2}{3}\right)&0&-\dfrac{c_{L,R}^\ast e}{\sqrt{2}c_Ws_W}\sin\epsilon\\
\mathbb{O}_3&0&-\dfrac{e}{c_Ws_W}\left(\dfrac{1}{2}+\dfrac{2s_w^2}{3}\right)&-\dfrac{c^\ast_{L,R}e}{\sqrt{2}c_Ws_W}\sin\epsilon\\
\mathbb{O}_3&-\dfrac{c_{L,R}e}{\sqrt{2}c_Ws_W}\sin\epsilon&-\dfrac{c_{L,R}e}{\sqrt{2}c_Ws_W}\sin\epsilon&-{2es_W}/{3c_W}
\end{pmatrix}
\end{split}
\end{align}
}}
and
{\footnotesize{
\begin{align}\label{eq: Z down MCHM}
\begin{split}
A_{NC}^{bL,R}&=
\begin{pmatrix}
\dfrac{e}{c_Ws_W}\left(-\dfrac{\delta^L}{2}+\dfrac{s_W^2}{3}\right)\mathbb{I}_3& \mathbb{O}_3^T\\
\mathbb{O}_3&\dfrac{e}{c_Ws_W}\left(-\dfrac{1}{2}+\dfrac{s_W^2}{3}\right)
\end{pmatrix}
\end{split}\,.
\end{align}
}}
For the charged currents mediated by the $W$ we have the interaction Lagrangian
\begin{equation}\label{eq: W MCHM}
\mathcal{L}\supset
W^+_\mu \overline{
\xi}_{\uparrow L,R}
\gamma^\mu
A_{CC}^{L,R}
\xi_{\downarrow L,R}\,,\quad\text{with}\quad
{\footnotesize{
A_{CC}^{L,R}=\dfrac{g}{\sqrt{2}}
\begin{pmatrix}
\delta^L\mathbb{I}_{3\times3}& \mathbb{O}_3^T\\
\mathbb{O}_3&\cos^2{\epsilon}/{2}\\
\mathbb{O}_3&\sin^2{\epsilon}/{2}\\
\mathbb{O}_3&-c_{L,R}^\ast\sin\epsilon
\end{pmatrix}
}}\,.
\end{equation}
In the above formulas, $\mathbb{O}_3 = (0, 0, 0)$.

\subsection{Perturbative expansion}\label{sec:pertexp}

In order to obtain simple analytic results, we diagonalise the mass matrices, presented in the previous section, in perturbation theory. We use as a perturbation parameter $\epsilon = v/f$. In the following we review in general terms the adopted procedure and then apply it to the down-type quark sector as an example, giving the explicit expressions for the matrices $U_{dL,R}$ which block-diagonalize the down-type mass matrix up to $\mathcal{O}(\epsilon^3)$ corrections. 
\bigskip

Given a general squared matrix $M(\lambda)$ with a parameter dependence, we can find its eigenvalues and eigenvectors at second order in $\lambda$ taking the following steps:

\begin{itemize}
\item Expand the matrix $M(\lambda)$ in $\lambda$ up to the second power;
\begin{equation}
M(\lambda)=M_{0}+\lambda M_{1}+\lambda^2 M_{2}+\mathcal{O}(\lambda^3)\,.
\end{equation}

\item Define the hermitian combinations (left and right) in powers of $\lambda$;
\begin{align}
\begin{split}
\text{Zeroth order:\quad}&H_{0L}=M_0M_0^\dagger\\
\text{First order:\quad}&H_{1L}=M_0M_1^\dagger+M_1M_0^\dagger\\
\text{Second order:\quad}&H_{2L}=M_0M_2^\dagger+M_2M_0^\dagger+M_1M_1^\dagger
\end{split}
\end{align}
(for the right combinations one just needs to replace $M_i\rightarrow M_i^\dagger$.)

\item Determine the zeroth order eigenvalues and eigenvectors, {\emph{i.e.}} $\lambda_0[i]$ and $\left|i_{L,R}\right>_0$, defined by the equation $H_{0L,R}\left|i_{L,R}\right>_0=\lambda_{0}[i]\left|i_{L,R}\right>_0$.

\item Use the second order expression for the eigenvectors
\begin{align}\label{eq: sum pert theory}
\begin{split}
\left|i_{L,R}\right>_2=&\left|i_{L,R}\right>_0+\lambda \sum_{k\neq i}^n\dfrac{{}_0\!\left<k_{L,R}\right|H_{1L,R}\left|i_{L,R}\right>_0}{\lambda_0[i]-\lambda_0[k]}\left|k_{L,R}\right>_0\\
&+\lambda^2\left(\sum_{k\neq i}^n\dfrac{{}_0\!\left<k_{L,R}\right|H_{2L,R}\left|i_{L,R}\right>_0}{\lambda_0[i]-\lambda_0[k]}\left|k_{L,R}\right>_0\right.\\
&+\sum_{k\neq i}^n\sum_{m\neq i}^n\dfrac{{}_0\!\left<k_{L,R}\right|H_{1L,R}\left|m_{L,R}\right>_{0}{}_0\!\left<m_{L,R}\right|H_{1L,R}\left|i_{L,R}\right>_0}{(\lambda_0[i]-\lambda_0[k])(\lambda_0[i]-\lambda_0[m])}\left|k_{L,R}\right>_0\\
&\left.-\dfrac{1}{2}\sum_{k\neq i}^n\left(\dfrac{{}_0\!\left<k_{L,R}\right|H_{1L,R}\left|i_{L,R}\right>_0}{\lambda_0[i]-\lambda_0[k]}\right)^2\left|i_{L,R}\right>_0\right)
\end{split}
\end{align}

\item In the sums in the Eq.(\ref{eq: sum pert theory}) it is implicitly assumed that we skip those indices $i\neq j$ for which $\lambda_0[i]=\lambda_0[j]$. By not summing these contributions we loose orthogonality between the zeroth order degenerate states at order $\lambda^2$. Therefore we must orthogonalize them, for instance through the known Gram-Schmidt procedure
\begin{equation}
\left|j\right>^{\mbox{new}}=\left|j\right>-\dfrac{\left<i\right|\left.j\right>}{\left<i\right|\left.i\right>}\left|i\right>\,.
\end{equation}
\end{itemize}
By following the procedure above we are able to find matrices $U_{L,R}=\left(\left|1_{L,R}\right>_2,\left|2_{L,R}\right>_2,\cdots\right)$ which are unitary up $\mathcal{O}(\lambda^2)$ and which (block) diagonalize $M_{\rm up}$ and $M_{\rm down}$. 

To make this procedure less abstract let us look to the simple scenario of the down sector. We start by identifying the expansion parameter $\lambda$ as $\sin 2\epsilon$. The relevant hermitian combinations take the form
\begin{align}
\begin{split}
\text{Zeroth order:}&\quad H_{0L}^d=
\begin{pmatrix}
0&0&0&0\\
0&0&0&0\\
0&0&f^2 y_{L4}^2&fM_4y_{L4}\\
0&0&fM_4y_{L4}&M_4^2
\end{pmatrix}\,,\quad H_{0R}^d=
\begin{pmatrix}
0&0&0&0\\
0&0&0&0\\
0&0&0&0\\
0&0&0&M_T^2
\end{pmatrix}\\
\text{First order:}&\quad H_{1L}^d=\mathbb{O}\,,\quad H_{0R}^d=\dfrac{fy_{4L}}{2}
\begin{pmatrix}
0&0&0&m^d_{\rm UV 31}\\
0&0&0&m^d_{\rm UV 32}\\
0&0&0&m^d_{\rm UV 33}\\
m^d_{\rm UV 31}&m^d_{\rm UV 32}&m^d_{\rm UV 33}&0
\end{pmatrix}\\
\text{Second order:}&\quad H_{2L}^d=
\begin{pmatrix}
&&&0\\
&m_{\rm UV}^d.m_{\rm UV}^{d\,T}&&0\\
&&&0\\
0&0&0&0
\end{pmatrix}\,,\quad H_{0R}^d=\begin{pmatrix}
&&&0\\
&m_{\rm UV}^{d\,T}.m_{\rm UV}^d&&0\\
&&&0\\
0&0&0&0
\end{pmatrix}
\end{split}
\end{align}
The next step is to find the eigensystem at zero order. In our scenario we have:
\begin{align}
\begin{split}
\text{Eigenvalues:}&\quad \lambda_0[1]=\lambda_0[2]=\lambda_0[3]=0\,,\quad \lambda_0[4]=M_T^2\,,\\
\text{Eigenvectors:}&\quad 
\left\{\begin{array}{l}
\left|1_L\right>_0=
\begin{pmatrix}
1\\0\\0\\0
\end{pmatrix}\,,\, 
\left|2_L\right>_0=
\begin{pmatrix}
0\\1\\0\\0
\end{pmatrix}\,,\,
\left|3_L\right>_0=
\begin{pmatrix}
0\\0\\ c_{\phi L}\\-s_{\phi L}
\end{pmatrix}\,,\,
\left|4_L\right>_0=
\begin{pmatrix}
0\\0\\ s_{\phi L}\\ c_{\phi L}
\end{pmatrix}\\
\left|i_R\right>_0=(\cdots \delta_{ij}\cdots)^T
\end{array}\right.
\end{split}
\end{align} 
There is an arbitrariness in the choice of the first three zero-order eigenvectors, due to the degeneracy of the light sector. As previously explained, contrarily to the usual perturbation theory, we shall not solve this degeneracy at first order before going to higher orders. Instead we keep this degeneracy to second order and use the perturbative expression for the eigenvectors presented in Eq.\eqref{eq: sum pert theory}. At the end of this stage, the second-order eigenvectors that shared the same eigenvalue at zero order are no longer orthogonal. We then do the last step, {\emph{i.e.}} the Gram-Schmidt orthogonalization procedure, leading to the unitary (up to $\mathcal{O}(s_{2\epsilon}^2)$) $4\times 4$ rotation matrices

\begin{equation}\label{eq:UdLpert}
\footnotesize
U_{dL}=
\begin{pmatrix}
1&0&0&s_{2\epsilon}^2s_{\phi L}\dfrac{m^2_{13}}{M_T^2}\\
0&1&0&s_{2\epsilon}^2s_{\phi L}\dfrac{m^2_{23}}{M_T^2}\\
-s_{2\epsilon}^2s_{\phi L}^2\dfrac{m^2_{13}}{M_T^2}&-s_{2\epsilon}^2s_{\phi L}^2\dfrac{m^2_{23}}{M_T^2}&c_{\phi L}\left(1-s_{2\epsilon}^2s_{\phi L}^2\dfrac{m^2_{33}}{M_T^2}\right)&s_{\phi L}\left(1+s_{2\epsilon}^2c_{\phi L}^2\dfrac{m^2_{33}}{M_T^2}\right)\\
-s_{2\epsilon}^2s_{\phi L}c_{\phi L}\dfrac{m^2_{13}}{M_T^2}&-s_{2\epsilon}^2s_{\phi L}c_{\phi L}\dfrac{m^2_{23}}{M_T^2}&-s_{\phi L}\left(1+s_{2\epsilon}^2c_{\phi L}^2\dfrac{m^2_{33}}{M_T^2}\right)&c_{\phi L}\left(1-s_{2\epsilon}^2s_{\phi L}^2\dfrac{m^2_{33}}{M_T^2}\right)
\end{pmatrix}\, ,
\end{equation}
\begin{equation*}
\footnotesize
U_{dR}=
\begin{pmatrix}
1-s_{2\epsilon}^2 s_{\phi L}^2\dfrac{m_{\rm UV 31}^{d2}}{8M_T^2}&-s_{2\epsilon}^2s_{\phi L}^2\dfrac{m^d_{\rm UV 31}m^d_{\rm UV 32}}{4M_T^2}&-s_{2\epsilon}^2s_{\phi L}^2\dfrac{m^d_{\rm UV 31}m^d_{\rm UV 33}}{4M_T^2}&s_{2\epsilon} s_{\phi L}\dfrac{m_{\rm UV 31}}{2M_T}\\
0&1-s_{2\epsilon}^2 s_{\phi L}^2\dfrac{m_{\rm UV 32}^{d2}}{8M_T^2}&-s_{2\epsilon}^2s_{\phi L}^2\dfrac{m^d_{\rm UV 32}m^d_{\rm UV 33}}{4M_T^2}&s_{2\epsilon} s_{\phi L}\dfrac{m_{\rm UV 32}}{2M_T}\\
0&0&1-s_{2\epsilon}^2 s_{\phi L}^2\dfrac{m_{\rm UV 33}^{d2}}{8M_T^2}&s_{2\epsilon} s_{\phi L}\dfrac{m_{\rm UV 33}}{2M_T}\\
-s_{2\epsilon} s_{\phi L} \dfrac{m_{\rm UV 31}^d}{2M_T}&-s_{2\epsilon} s_{\phi L} \dfrac{m_{\rm UV 32}^d}{2M_T}&-s_{2\epsilon} s_{\phi L} \dfrac{m_{\rm UV 33}^d}{2M_T}&1-s_{2\epsilon}^2  s_{\phi L}^2 \dfrac{m_{33}^2}{8M_T^2}
\end{pmatrix}\, ,
\end{equation*}
where in the above we use the shorthand $m^2_{\alpha\beta}=m^d_{\rm UV \alpha i}m^d_{\rm UV \beta i}$. 

The rotations $U_{tL,R}$ which block-diagonalize the up-type mass matrix Eq.(\ref{eq: Mup appendix}) are obtained analogously, but the expressions are larger so that we do not give them, here.

\section{A variation: fully composite right-handed top} \label{app:comptR}
A minimal variation to the scenario considered here is to assume that the right-handed top is a massless composite state of the dynamics. This assumption leads to changes in the up-sector  mass, Yukawa  matrices and gauge interactions. The Lagrangian, analogous to Eq.\eqref{eq:Lag tot PC}, now reads:
\begin{align}
\begin{split}
\mathcal{L}_{comp}=&i\overline{Q}_{L,R}\left(D\s+E\s\right) Q_{L,R}+i\overline{\tilde{T}}_{L,R} D\s\tilde{T}_{L,R}-M_4\left(\overline{Q}_L Q_R+\overline{Q}_R Q_L\right)\\
&-M_1\left(\overline{\tilde{T}}_L\tilde{T}_R+\overline{\tilde{T}}_R\tilde{T}_L\right)+ic_L \overline{Q}_L^i\gamma^\mu d^i_\mu \tilde{T}_L+ic_R \overline{Q}_R^i\gamma^\mu d^i_\mu \tilde{T}_R+ic_t \overline{Q}_R^i\gamma^\mu d^i_\mu t_R+\text{h.c.}\\
-\mathcal{L}_{mix}=&y_{L4,1}f \overline{q}_{3L}^5 U\psi_R+y_{Lt}f \overline{q}_{3L}^5Ut_R+\text{h.c.}\\
=&y_{L4}f \left(\overline{b}_L B_R+c^2_{\theta/2}\overline{t}_L T_R+s_{\theta/2}^2\overline{t}_L X_{2/3R}\right) -\frac{y_{L1}f}{\sqrt{2}}s_\theta \overline{t}_L\tilde{T}_R-\frac{y_{Lt}f}{\sqrt{2}}s_\theta \overline{t}_L\tilde{t}_R +h.c.
\end{split}
\end{align}
The mass matrix reads:
\begin{equation}
\footnotesize{
M_{\rm up}=
\begin{pmatrix}
\tilde{m}[\epsilon]_{11}&\tilde{m}[\epsilon]_{12}&\tilde{m}[\epsilon]_{13}&0&0&0\\
\tilde{m}[\epsilon]_{21}&\tilde{m}[\epsilon]_{22}&\tilde{m}[\epsilon]_{23}&0&0&0\\
\tilde{m}[\epsilon]_{31}&\tilde{m}[\epsilon]_{32}&\tilde{m}[\epsilon]_{33} - f\dfrac{y_{Lt}}{\sqrt{2}}\sin\epsilon&fy_{L4}\cos^2\dfrac{\epsilon}{2}&fy_{L4}\sin^2\dfrac{\epsilon}{2}&-f\dfrac{y_{L1}}{\sqrt{2}}\sin\epsilon\\
0&0&0&M_4&0&0\\
0&0&0&0&M_4&0\\
0&0&0&0&0&M_1
\end{pmatrix}
}
\end{equation}
The diagonalisation of this mass matrix will generate corrections to the SM quark couplings similar to the ones in the case of an elementary $t_R$. The masses for the top and the heavy partners are:
\begin{eqnarray}
 m_t = \big|\frac{M_{4}  f
y_{Lt}}{2 \sqrt{2} \sqrt{ M_{4}^2 + f^2 y_{L4}^2}}\big| s_{2\epsilon}\,, \quad M_{T}=\sqrt{ M_{4}^2 + f^2 y_{L4}^2}\,, \quad  M_{X_{2/3}} = M_4\,, \quad  M_{\tilde T} = M_{1} \end{eqnarray}
Combining the contributions from differentiating of $M_{up}$  and from the $d_\mu^4$-term of  $\mathcal{L}_{comp}$,  the Yukawa matrix is:
\begin{equation} \footnotesize{
 Y_{up}=
\begin{pmatrix} \tilde{y}[\epsilon]_{2\times2} & \tilde{y}[\epsilon]_{2}^T & \mathbb{O}_{2}^T & \mathbb{O}_{2}^T & \mathbb{O}_{2}^T \\
\tilde{y}[\epsilon]_{2} & \tilde{y}[\epsilon]_{33} + \left(c_t y_{Lt} - \frac{ y_{Lt}}{\sqrt{2}}\right)c_\epsilon &\left(\frac{c_R^\ast y_{L1}}{\sqrt{2}}+\frac{c_t^\ast y_{Lt}}{\sqrt{2}} - \frac{y_{L4}}{2}\right)s_\epsilon &-\left(\frac{c_R^\ast y_{L1}}{\sqrt{2}}+ \frac{c_t^\ast y_{Lt}}{\sqrt{2}} -\frac{y_{L4}}{2}\right)s_\epsilon& \left(c_Ry_{L4}-\frac{ y_{L1}}{\sqrt{2}}\right)c_\epsilon\\ \mathbb{O}_{2} &
c_t \frac{M_4}{f} &0&0&-\frac{c_L M_1-c_R M_4}{f}\\ \mathbb{O}_{2}&
-c_t \frac{M_4}{f} &0&0&-\frac{c_R M_4-c_LM_1}{f}\\ \mathbb{O}_{2} &
0 &-\frac{-c_L^\ast M_4+c_R^\ast M_1}{f}&-\frac{c_L^\ast
M_4-c_R^\ast M_1}{f}&0
\end{pmatrix}}\,
\end{equation}
Rotating into the heavy quark eigenstates, we find that  the factorization pattern displayed in the  partial composite case continue to hold, and the additional $c_t$ term in the $\mathcal{L}_{comp}$ will give rise to  an $\mathcal{O}(1)$ correction in the Yukawa interaction:
\begin{eqnarray} 
&&  m_U \simeq  \frac{s_{2\epsilon}}{2}\ m^u_{\rm UV} \mp
m_t\Pi \,, \quad y_u \simeq 
\dfrac{m_U}{fs_{2\epsilon}/2}\left(1-\frac{1}{2}s_{2\epsilon}^2\right)+
c_t \left(y_{Lt} -y_{L4}\right) c_{\phi L} \Pi +B_u \,,\nonumber \\ 
&& \mbox{where} ~~  \Pi=\small{\left(\begin{array}{ccc}0&0&0\\0&0&0\\0&0&1\end{array}\right)}\,, \quad B_u\sim\frac{\Sigma_u}{M_*^2} \,,  \quad \mbox{and} ~~  c_{\phi L} = \frac{M_*}{\sqrt{M_*^2 + f^2 y_{L4}^2}}\,.
\end{eqnarray}

We continue to discuss the deviations in the gauge interaction. When the $t_R$ is fully composite, the right-handed $Z$ and $W$ currents will be further corrected:
\begin{align}
\begin{split}
\mathcal{L}\supset  W^+_\mu \overline{\xi_{\uparrow R}}\gamma^\mu A_{CC}^{tR} \xi_{\downarrow R} + Z_\mu \overline{ \xi_{\uparrow R}}\gamma^\mu A_{NC}^{tR} \xi_{\uparrow R}
\end{split}
\end{align}
where the matrices $A_{CC}^{tR}$ and $ A_{NC}^{tR}$ in the flavour basis are:
\begin{equation}\footnotesize{
A_{CC}^{tR}=\frac{g}{\sqrt{2}}
\begin{pmatrix}
\mathbb{0}_{3\times2} & \mathbb{0}_{2}^T  \\ \mathbb{0}_{3} & - c_t^\ast \sin \epsilon\\ 
\mathbb{0}_{3}  &\cos^2{\epsilon}/{2}\\
\mathbb{0}_{3}  &\sin^2{\epsilon}/{2}\\
\mathbb{0}_{3}  &-c_{R}^\ast\sin\epsilon
\end{pmatrix}}\,
\end{equation}
{\footnotesize{
\begin{align}
\begin{split}
A_{NC}^{tR}=
\begin{pmatrix}  -\frac{2es_W}{3c_W} \mathbb{I}_2  & \mathbb{0}_2^T & \mathbb{0}_2^T &  \mathbb{0}_2^T & \mathbb{0}_2^T \\
\mathbb{0}_2 & -\frac{2e s_W}{3 c_W} &-\frac{c_{t}^\ast e}{\sqrt{2}c_Ws_W}\sin\epsilon&-\frac{c_{t}^\ast e}{\sqrt{2}c_Ws_W}\sin\epsilon&0\\
\mathbb{0}_2 & -\frac{c_{t}e}{\sqrt{2}c_Ws_W} \sin
\epsilon&\frac{e}{c_Ws_W}\left(\frac{1}{2}-\frac{2s_w^2}{3}\right)&0&
-\frac{c_{R}^\ast e}{\sqrt{2}c_Ws_W}\sin\epsilon\\
\mathbb{0}_2 & -\frac{c_{t}e}{\sqrt{2}c_Ws_W} \sin \epsilon
&0&-\frac{e}{c_Ws_W}\left(\frac{1}{2}+\frac{2s_w^2}{3}\right)& 
-\frac{c^\ast_{R}e}{\sqrt{2}c_Ws_W}\sin\epsilon \\
\mathbb{0}_2 & 0&-\frac{c_{R}e}{\sqrt{2}c_Ws_W}\sin\epsilon&
-\frac{c_{R}e}{\sqrt{2}c_Ws_W}\sin\epsilon& -\frac{2es_W}{3c_W}
\end{pmatrix}
\end{split}
\end{align}}}
The deviations in charged and neutral right-hand currents should be calculated by transforming to the block diagonal basis, and are of the same order as in the partial composite case. The main difference is that a fully composite $t_R$ couples directly to massive resonances of the underlying dynamics, therefore higher order contributions may arise and differ from the partial composite case in the up-sector. Let us first consider the composite top partners coupling to a singlet scalar $\Phi$ as dictated by Eq.(\ref{eq: Phi interaction}). The flavour interaction for the SM up-type quarks brought by the partial compositeness effect is:
\begin{equation}
\mathcal{L}_S\simeq
\Phi\left(\begin{array}{ccc}\bar{u}_L&\bar{c}_L&\bar{t}_L\end{array}\right)\cdot\left(\begin{array}{ccc}
 0 & 0 & 0 \\
 0 & 0 & 0 \\
 g_{B}s_{\phi L}^2c_{\phi L} \frac{ m_{c} }{M_*} &  g_{B}s_{\phi L}^2c_{\phi L}\frac{ m_{c} }{M_*} & - g_{B}s_{\phi L}^2\frac{ m_{t}}{M_*}
\end{array}\right)\cdot
\left(\begin{array}{c}u_R\\c_R\\t_R\end{array}\right)\,+h.c.
\end{equation}
with the notation $s_{\phi L}= f y_{L4}/ \sqrt{M_*^2+f^2 y_{L4}^2}$. A difference arises with respect to Eq.(\ref{eq:Yukawa Phi up}) since there is no $t_R$ mixing before EWSB and thus $s_{\phi R}=0$ in this scenario. When we  set $m_{\Phi}=g_B f$ and integrate out the scalar resonance,  the effective Lagrangian for the dimension-6 operator is,
\begin{equation}
\mathcal{L}_S\simeq \left(1-2 c_{\phi L}\right)^2\frac{s_{\phi L}^4}{f^2}{\left(\frac{m_c}{m_t}\right)}^4{\left(\frac{m_t}{M_*}\right)}^2\mathcal{Q}_4^{uc}\simeq \frac{10^{-10}}{\mbox{ TeV}^2} \left( \frac{1\ \mbox{TeV}}{M_\ast} \right)^2\mathcal{Q}_4^{uc}\,.
\end{equation}
The coefficient $C_{4}^{u c}$ is  well below the experimental bound. A larger difference comes from  the $t_R$ coupling to a vector resonance $V_\mu$ in a $\SO(4)$ singlet since the chiral property permits the following interaction,
\begin{equation}
\mathcal{L}_V=V_\mu(g_B \bar{Q}_{L}\gamma^\mu Q_{L}+ g_S
\bar{\tilde{T}}_{L}\gamma^\mu\tilde{T}_{L})+(L\rightarrow
R)+ g_S^\prime
\bar{t}_{R}\gamma^\mu t_{R}+\frac{1}{2}m_V^2V_\mu V^\mu\,.
\end{equation}
Assuming  that $y_{L1} = y_{L4}$, $g_S = g_B $ and $ M_* = M_4 = M_1$, and  diagonalizing the heavy quark mass, the flavour interaction for the SM up-type  quarks is:
{ \footnotesize{\begin{eqnarray}
&& \mathcal{L}_V\simeq
V_\mu\left(\begin{array}{ccc}\bar{u}_L &\bar{c}_L &
\bar{t}_L \end{array}\right) \gamma^\mu
\cdot\left(\begin{array}{ccc}
0&0& -g_B s_{\phi L}^2 c_{\phi L}^2\frac{m_t m_c}{M_*^2}\\
0&0& -g_B s_{\phi L}^2 c_{\phi L}^2\frac{m_t m_c}{M_*^2}\\
-g_B s_{\phi L}^2  c_{\phi L}^2\frac{m_t m_c}{M_*^2}& -g_Bs_{\phi L}^2 c_{\phi L}^2\frac{m_t m_c}{M_*^2}& g_B  s_{\phi L}^2 + 2 g_B s_{\phi L}^2  c_{\phi L}^2 \frac{m_t^2}{M_*^2}\end{array}\right)\cdot
\left(\begin{array}{c}u_L \\c_L \\t_L \end{array}\right)  \nonumber  \\
&& + ~ V_\mu\left(\begin{array}{ccc}\bar{u}_R  &\bar{c}_R &
\bar{t}_R \end{array}\right) \gamma^\mu
\cdot\left(\begin{array}{ccc}
g_B s_{\phi L}^2 c_{\phi L}^2\frac{ m_c^2}{M_*^2}&g_B s_{\phi L}^2 c_{\phi L}^2\frac{m_c^2}{M_*^2}& -g_B s_{\phi L}^2 c_{\phi L}\frac{m_t m_c}{M_*^2}\\
g_B s_{\phi L}^2 c_{\phi L}^2\frac{m_c^2}{M_*^2}&g_B s_{\phi L}^2 c_{\phi L}^2\frac{m_c^2}{M_*^2}& -g_B s_{\phi L}^2 c_{\phi L}\frac{m_t m_c}{M_*^2}\\ -g_B s_{\phi L}^2c_{\phi L}\frac{m_t m_c}{M_*^2}& -g_Bs_{\phi L}^2c_{\phi L}\frac{m_t m_c}{M_*^2}& g_S^\prime +\left(g_B -g_S^\prime\right) s_{\phi L}^2 \frac{ m_t^2}{M_*^2}\end{array}\right)\cdot
\left(\begin{array}{c}u_R \\c_R \\t_R \end{array}\right) \end{eqnarray}}}
We further rotate from the flavour basis into the mass  basis and get the  Lagrangian:
\begin{eqnarray}
&& \mathcal{L}_V \simeq \left( g_B s_{\phi L}^2
{\left(\frac{m_c}{m_t}\right)}^2 - 2 g_B s_{\phi L}^2 c_{\phi L}^2
{\left(\frac{m_c}{M_*}\right)}^2 \right)\,V_\mu \, \bar{u}_L
\gamma^\mu c_L \nonumber \\ && +\left(g_S^\prime
{\left(\frac{m_c}{m_t}\right)}^2 + \left(g_B\left(1-2 c_{\phi
L}\right)^2 - g_S^\prime\right) s_{\phi L}^2
{\left(\frac{m_c}{M_*}\right)}^2 \right) \,V_\mu\, \bar{u}_R
\gamma^\mu c_R + h.c.
\end{eqnarray}
Integrating out the heavy resonance with its  mass set to be  $m_V
= g_B f$, we find that the dimension-6  operators have the
coefficients:
\begin{eqnarray}&& \mathcal{Q}_1^{u c}\,: 
\quad  \frac{1}{f^2}{\left(s_{\phi
L}^2\left(\frac{m_c}{m_t}\right)^2-2 s_{\phi L}^2 c_{\phi L}^2
\left(\frac{m_c}{M_*}\right)^2 \right)}^2  \nonumber \\
&& \tilde{\mathcal{Q}}_1^{u c} \,: \quad
\frac{1}{f^2}\left(\frac{g_S^{\prime}}{g_B}{\left(\frac{m_c}{m_t}\right)}^2
+\left(\left(1-2 c_{\phi
L}\right)^2-\frac{g_S^{\prime}}{g_B}\right) s_{\phi L}^2
{\left(\frac{m_c}{M_*}\right)}^2 \right)^2\,.
\end{eqnarray} 
Therefore the Wilson coefficients $C_1^{u c} $ and $\tilde{C}_1^{ u c}$ will be of order of  $\sim 10^{-9}/\mbox{TeV}^2$, which are  below the experimental bound
(see Ref.~\cite{ Calibbi:2012at} for  $D^0-\bar{D}^0$ constraints on the real and  imaginary parts
of  $C_1^{uc}$ or  $\tilde{C}_1^{uc}$, which are of  the same magnitude).

\end{appendix}

\end{document}